%% file: paper.tex
\documentclass[sigconf]{acmart} 
\renewcommand\footnotetextcopyrightpermission[1]{} %

\usepackage[utf8]{inputenc}
\usepackage{arcs}
\usepackage{color}
\usepackage{graphicx}
\usepackage{latexsym}
\usepackage{psfrag}
\usepackage{tikz}
\usepackage{color, colortbl}

\usepackage{algorithm}
\usepackage{algorithmicx}
\usepackage{algcompatible}
\usepackage{soul}
\usepackage{caption}
\usepackage{subcaption}
\usepackage{float}
\usepackage{pifont}
\usetikzlibrary{arrows,shapes}
\usepackage{multirow}
\usepackage{url}
\usepackage{xspace}
\usepackage{hyperref}
\usepackage{mdframed}
\usepackage{pgf-umlsd}
\usepackage{tikz}
\usepackage{amsthm}
\usepackage{booktabs}
\usetikzlibrary{positioning,arrows.meta,fit,backgrounds,calc}

\newcommand{\kETAL}    {{\em et~al.}}

\newcommand{\setup}{\mathsf{ABE.Setup}}

\newcommand{\encrypt}{\mathsf{ABE.Encrypt}}
\newcommand{\decrypt}{\mathsf{ABE.Decrypt}}

\newcommand{\fregister}{\mathcal{F}_{\mathsf{register}}}

\newcommand{\sid}{\mathit{sid}}

\newcommand{\spid}{\mathit{spid}}

\newcommand{\fresp}{\mathcal{F}_{\mathsf{response}}}
\newcommand{\F}[1]{\mathcal{F}_{\mathsf{#1}}}
\newcommand{\fsig}{\mathcal{F}_{\mathsf{sig}}}
\newcommand{\ftee}{\mathcal{F}_{\mathsf{tee}}}
\newcommand{\fsmt}{\mathcal{F}_{\mathsf{smt}}}

\newcommand{\fdctc}{\mathcal{F}_{\mathsf{LATTEO}}}

\newcommand{\symmetable}{\mathit{symmETable}}
\newcommand{\ABEetable}{\mathit{ABEeTable}}
\newcommand{\symmktable}{\mathit{symmKTable}}
\newcommand{\spidapptable}{\mathit{spAppTable}}

\newcommand{\stable}{\mathit{sTable}}
\newcommand{\abesktable}{\mathit{ABEskTable}}

\newcommand{\etable}{\mathit{eTable}}

\newcommand{\symmkeygen}{\mathsf{symmKeygen}}
\newcommand{\symmenc}{\mathsf{symmEnc}}
\newcommand{\symmdec}{\mathsf{symmDec}}

\newcommand{\abesetup}{\mathsf{abeSetup}}
\newcommand{\abekeygen}{\mathsf{abeKeyGen}}
\newcommand{\abeenc}{\mathsf{abeEnc}}
\newcommand{\abedec}{\mathsf{abeDec}}

\newcommand{\KeyGen}{\mathsf{KeyGen}}

\newcommand{\party}{\mathcal{P}}
\newcommand{\getpk}{\mathrm{getpk}}
\newcommand{\prog}{\mathrm{prog}}
\newcommand{\reg}{\mathrm{reg}}
\newcommand{\inp}{\mathrm{inp}}
\newcommand{\mem}{\mathrm{mem}}
\newcommand{\outp}{\mathrm{outp}}
\newcommand{\eid}{eid}

\makeatletter
\newcommand{\NoEndLine}[1]{%
  \setbox\@tempboxa\hbox{#1}%
  \unhbox\@tempboxa\unskip\ignorespaces}
\makeatother

\newcommand{\simu}{\mathcal{S}}
\newcommand{\env}{\mathcal{Z}}
\newcommand{\adv}{\mathcal{A}}

\newcommand{\gzero}{\mathbf{Game~0}}
\newcommand{\gone}{\mathbf{Game~1}}
\newcommand{\gtwo}{\mathbf{Game~2}}

\newcommand{\execgzero}{\mathsf{Exec_{\mathsf{Game0},\mathcal{Z}}}}
\newcommand{\execgone}{\mathsf{Exec_{\mathsf{Game1},\mathcal{Z}}}}
\newcommand{\execgtwo}{\mathsf{Exec_{\mathsf{Game2},\mathcal{Z}}}}

\definecolor{SeaGreen}{rgb}{0.18, 0.55, 0.34}
\usepackage [
n ,
advantage ,
operators,
sets,
adversary ,
landau ,
probability ,
notions, 
logic,
ff,
mm,
primitives,
events,
complexity,
asymptotics ,
keys
]{cryptocode}

\floatname{algorithm}{Protocol}

\newcommand{\eg}{{\it e.g., }}
\newcommand{\ie}{{\it i.e., }}
\newtheorem{lemma}{Lemma}[section]
\newtheorem{theorem}{Theorem}[section]
\newtheorem{definition}{Definition}[section]

\def\blackslug{\hbox{\hskip 1pt \vrule width 4pt height 8pt depth 1.5pt
  \hskip 1pt}}
\def\QED{\quad\blackslug\lower 8.5pt\null}
\def\BibTeX{{\rm B\kern-.05em{\sc i\kern-.025em b}\kern-.08em
    T\kern-.1667em\lower.7ex\hbox{E}\kern-.125emX}}
    
\makeatletter
\def\ps@headings{%
\def\@oddhead{\mbox{}\scriptsize\rightmark \hfil \thepage}%
\def\@evenhead{\scriptsize\thepage \hfil \leftmark\mbox{}}%
\def\@oddfoot{}%
\def\@evenfoot{}}
\makeatother
\def\BibTeX{{\rm B\kern-.05em{\sc i\kern-.025em b}\kern-.08em
    T\kern-.1667em\lower.7ex\hbox{E}\kern-.125emX}}
\makeatletter
\@addtoreset{equation}{section}
\makeatother

\newcommand{\graybox}[1]{%
    \colorbox{gray!30}{%
        \parbox{\dimexpr\linewidth-2\fboxsep}{%
            \centering %
            #1
        }%
    }%
}

\newcommand{\sysname} {{\em LATTEO}\xspace}

\begin{document}
\fancyhead{}
\title{LATTEO: A Framework to Support Learning Asynchronously Tempered with Trusted Execution and Obfuscation}

\authors{
    \author{Abhinav Kumar}
    \affiliation{%
      \institution{Saint Louis University}
      \city{St. Louis}
      \country{U.S.}      
      }
    \email{abhinav.kumar@slu.edu}

    \author{George Torres}
    \affiliation{%
      \institution{New Mexico State University}
      \city{Las Cruces}
      \country{U.S.}      
      }
    \email{gtorresz@nmsu.edu}
    
    \author{Noah Guzinski}
    \affiliation{%
      \institution{Saint Louis University}
      \city{St. Louis}
      \country{U.S.}      
      }
    \email{noah.guzinski@slu.edu}

    \author{Gaurav Panwar}
    \affiliation{%
      \institution{New Mexico State University}
      \city{Las Cruces}
      \country{U.S.}     
      }
    \email{gpanwar@nmsu.edu}

    \author{Reza Tourani}
    \affiliation{%
      \institution{Saint Louis University}
      \city{St. Louis}
      \country{U.S.}      
      }
    \email{reza.tourani@slu.edu}
    
    \author{Satyajayant Misra}
    \affiliation{%
      \institution{New Mexico State University}
      \city{Las Cruces}
      \country{U.S.}      
      }
    \email{misra@nmsu.edu}
    
    \author{\mbox{Marcin Spoczynski}}
    \affiliation{%
      \institution{Intel Labs}
      \city{Portland}
      \country{U.S.}      
      }  
      \email{marcin.spoczynski@intel.com}

    \author{Mona Vij}
    \affiliation{%
      \institution{Intel Labs}
      \city{Portland}
      \country{U.S.}       
      }
      \email{mona.vij@intel.com}   
      
      \author{Nageen Himayat}
    \affiliation{%
      \institution{Intel Labs}
      \city{Portland}
      \country{U.S.}       
      }
      \email{nageen.himayat@intel.com}        
    \renewcommand{\shortauthors}{A. Kumar \kETAL}
}

\begin{abstract}
The privacy vulnerabilities of the federated learning (FL) paradigm, primarily caused by gradient leakage, have prompted the development of various defensive measures. Nonetheless, these solutions have predominantly been crafted for and assessed in the context of synchronous FL systems, with minimal focus on asynchronous FL. This gap arises in part due to the unique challenges posed by the asynchronous setting, such as the lack of coordinated updates, increased variability in client participation, and the potential for more severe privacy risks. These concerns have stymied the adoption of asynchronous FL.
In this work, we first demonstrate the privacy vulnerabilities of asynchronous FL through a novel data reconstruction attack that exploits gradient updates to recover sensitive client data. To address these vulnerabilities, we propose a privacy-preserving framework that combines a gradient obfuscation mechanism with Trusted Execution Environments (TEEs) for secure asynchronous FL aggregation at the network edge. To overcome the limitations of conventional enclave attestation, we introduce a novel data-centric attestation mechanism based on Multi-Authority Attribute-Based Encryption. This mechanism enables clients to implicitly verify TEE-based aggregation services, effectively handle on-demand client participation, and scale seamlessly with an increasing number of asynchronous connections.
Our gradient obfuscation mechanism reduces the structural similarity index of data reconstruction by {\em 85\%} and increases reconstruction error by {\em 400\%}, while our framework improves attestation efficiency by lowering average latency by up to {\em 1500\%} compared to RA-TLS, without additional overhead.

\end{abstract}
\maketitle
\pagestyle{plain} 
\input{sec01}

\input{sec03}

\input{sec04}
\input{sec05}

\input{sec07}
\input{sec08}

\input{sec02}

\section{Conclusions}
\label{sec09}
In this work, we analyzed four attacks aimed at compromising clients' privacy in an asynchronous federated learning setting. To mitigate these attacks we proposed \sysname, comprised of our GOOD training and our TEE-mesh based system design. We show that GOOD significantly reduces the effectiveness of reconstruction attacks while preserving the model utility. We also show that our system design provides improved security for clients at minimal costs to both the server and the clients. Our system design also allows for vendor-agnostic client-side attestation, which makes the solution scalable and promotes trusted hardware heterogeneity.

\balance
\bibliographystyle{ACM-Reference-Format}
\bibliography{paper}

\appendix

\input{full-formalsecurity}

\input{appendix}

\end{document}

%% file: sec01.tex
\section{Introduction}
\label{sec01}
The increasing popularity of computationally intensive services, such as machine learning (ML), automated driving, and augmented reality (AR), is increasing demand for computational resources with low latency, high reliability, and strong security \& privacy guarantees. 
This has led to the emergence of edge and fog computing paradigms (e.g., pervasive edge computing) to allow for the delegation of these services to capable machines located at the network's edge. This addresses the applications' needs in a distributed and scalable manner~\cite{edgeCom}, while providing greater availability and reliability and lower latency to users than deployments on the cloud. 
\begin{figure}[t]
\centering
  \includegraphics[width=\columnwidth]{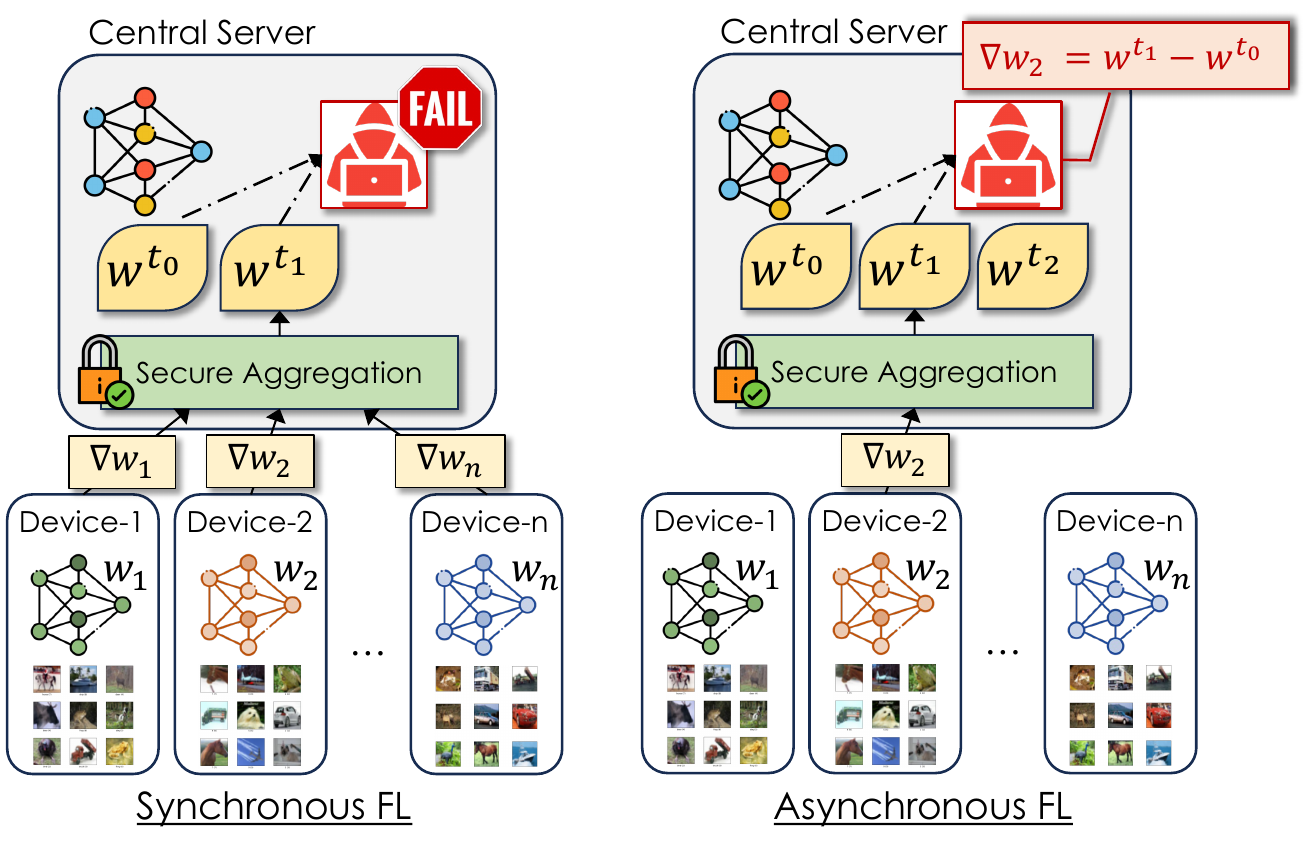} 
  \vspace{-0.3in}
  \caption{Asynchronous aggregation mechanism, although useful, leads to privacy vulnerability in asynchronous FL.}
  \label{fig:Introduction} 
  \vspace{-0.2in}
\end{figure}

Scalability of large edge ML applications have received a shot-in-the-arm with the adoption of distributed ML techniques, which can collaboratively leverage available resources of varying computational capabilities and operate at dynamic time scales.
However, the democratized and heterogeneous nature of the edge computing environment exacerbates some security flaws in the distributed models.  
Distributed learning models generally make use of either centralized training data or decentralized training data. The decentralized approach is more desirable for the sake of scalability and continuous learning for applications.  
Popular decentralized methods include Gossip Learning~\cite{Gossip} and Federated Learning (FL)~\cite{FederatedOpt}. Of the two, FL tends to be more efficient with an aggregator incorporating training data from various clients. 
FL can be done either synchronously or asynchronously~\cite{Caozhazha2024SRFL,FanLiuGon2022AFLGuard} as illustrated in Figure~\ref{fig:Introduction}. 

In a synchronous model, an aggregator will wait for all its clients to send in their weights before performing the aggregation. 
This method can provide some privacy to individual clients, as each model update reflects the sum of all the clients' weights.  
In a dynamic environment, such as the network edge, where clients may freely join and leave the system, this model is difficult to use. The aggregator may be left hanging on a client that has left or exclude clients that joined later. At the pervasive edge, an asynchronous model is more apt---the client’s weights will be aggregated as they are received or at regular time periods. While this model provides improved dynamicity, it loses some of the privacy provided by the synchronous model, as the model is updated upon receipt of a subset of clients' feedback instead of when all clients' feedback is received. As shown in Figure~\ref{fig:Introduction}, in this case the updates to the global model directly reflect incorporation of a client's weights. This allows a snooper (including client) that receives the server's iterative updates to reconstruct the updating clients' models by comparing the current and previous server updates. %

There are various works showcasing the vulnerability of both synchronous and asynchronous FL models to membership interference attacks and reconstruction attacks~\cite{nasshohou2019comprehensive,zhuLiGu2024evaluating,shostrmarc2017membership,yanGeXia2023using, BoeDziSchu2023when, ZhaShaElk2024Large-scale}. Several follow-on methods have been proposed to protect against said attacks, but most focus on synchronous FL~\cite{ShoRezShm2015Privacy,weilima2023personalized,Hewancai2024clustered,HuGuoGon2023Federated,xuliliu2020verifynet,NaHyeJun2022Closing,liuligao2023privacy-encoded,Caozhazha2024SRFL}. The differences between asynchronous and synchronous FL mean that these solutions cannot be properly mapped onto asynchronous FL. We  also present novel attack vectors we discovered, which are not addressed by these solutions. 

Some of these attack vectors are also made possible due to the inherent untrustworthiness of the aggregator's host machine. To address this lack of trust Trusted Execution Environment (TEE)~\cite{IEECiteforSGX, TEECite} have emerged. TEEs, with their controlled enclaves, contain a verification mechanism to ensure that only the program designated to handle a data runs inside, thus preserving the privacy of the client's data. An example of a TEE is Intel's SGX~\cite{IEECiteforSGX} which uses specialized hardware to create secure enclaves. 

\noindent
\textbf{Our Framework:}
To address these outstanding concerns, we present a framework for \emph{Learning Asynchronously, Tempered with Trusted Execution and Obfuscation}--referred to as \textit{\textbf{LATTEO}}. The \sysname framework addresses the privacy and scalability challenges in asynchronous federated learning (AsyncFL) through a three-component architecture. 
The first component is a \textit{\textbf{novel gradient obfuscation mechanism}} to safeguard client data privacy by entangling local gradient updates with a private orthogonal distribution in the latent space. Unlike differential privacy, which can degrade model utility, our gradient obfuscation scheme preserves the global model's utility with negligible computational overhead.
The second component is a \textit{\textbf{secure aggregation service mesh}} that employs confidential computing principles to enable the agile deployment of trusted aggregation services at the network edge. This is crucial in the edge-cloud ecosystem, where service providers delegate AsyncFL aggregation tasks to edge servers closer to clients, reducing latency and enhancing efficiency. To avoid constant reliance on cloud-based service providers, \sysname establishes a hierarchical trusted coordinator enclave structure, which is vendor-agnostic, and transfers the credentialing authority for attesting local aggregation enclaves and for provisioning credentials.
The vendor-agnostic design enables devices with different TEEs (e.g., Intel, Nvidia) to operate seamlessly in the ecosystem.    

Finally, \sysname framework features a novel \textit{\textbf{data-centric attestation mechanism}}, which provides a vendor-agnostic solution, enabling clients to implicitly verify aggregation services, using enclaves from multiple vendors in the same ecosystem. At its core, this mechanism leverages Multi-Authority Attribute-Based Encryption (MABE) to support asynchronous communication, aligning with the asynchronous nature of AsyncFL. This approach ensures scalability and eliminates reliance on hardware-specific attestation, outperforming conventional remote attestation methods.

We prototyped \sysname~and evaluated it with multiple neural network architectures and datasets. Results demonstrate that our gradient obfuscation mechanism preserves model utility, within {\em 1\%} of vanilla AsyncFL, while significantly reducing data reconstruction risks--lowering similarity metrics, such as structural similarity index by {\em 85\%} and increasing reconstruction error by {\em 400\%}. The system also scales seamlessly under high client loads, reducing average client latency by up to {\em 15x} compared to conventional RA-TLS.

The novel {\bf contributions} of \sysname are as follows: 

    \noindent {\bf (i)}
    We perform a thorough evaluation of AsyncFed by utilizing adversaries performing reconstruction attack with varying levels of prior knowledge. We do so by extending existing attack methodologies to fit the unique vulnerabilities that exist in AsyncFL, demonstrating privacy risks in AsyncFL systems.

    \noindent {\bf (ii)} We propose a gradient obfuscation mechanism to protects client privacy by entangling local gradient updates with an orthogonal distribution in the latent space. This entanglement is performed using a carefully designed loss function, ensuring that the gradients remain indistinguishable while maintaining model performance. 
    
    \noindent {\bf (iii)} We propose a scalable framework for the secure deployment of aggregation services at the network edge. \sysname incorporates a hierarchical enclave structure for trusted aggregation and a data-centric, vendor-agnostic attestation mechanism that provides scalable and implicit client verification, addressing limitations of conventional remote attestation methods. This attestation mechanism is generic and can be used widely in secure distributed systems.

    \noindent {\bf (iv)} We systematically evaluate \sysname using major benchmark datasets of varying complexity across two neural network architectures. We also assess scalability of the proposed hierarchical enclave structure and implicit attestation mechanism, comparing its operation cost to that of the conventional remote attestation-trasport layer security (RA-TLS) attestation scheme.

The rest of this paper is organized as follows. In Section~\ref{sec03} we discuss the threat model and elaborate on our gradient reconstruction attack. We then overview \sysname in Section~\ref{sec04}, and elaborate on its detailed design in Section~\ref{sec05}. Using the Universal Composability Framework, we prove \sysname's security in Section~\ref{formalsec}. Section~\ref{sec08} details the implementation scope, experimental setup, and analysis. Finally, we review the related work in Section~\ref{sec02}, and conclude our work in Section~\ref{sec09}.

%% file: sec03.tex
\section{Privacy Risks in Asynchronous FL}
\label{sec03} 
The existing literature has extensively investigated FL's privacy leakage risks, particularly those arising from the exposure of gradient updates during the model training~\cite{picromveg2023Perfectly,pasfraate2022Eluding,ZhaShaElk2024Large-scale,melsondec2019exploiting}. While current aggregation techniques effectively mitigate such privacy leakages in SyncFL~\cite{NaHyeJun2022Closing,liuligao2023privacy-encoded,weilima2023personalized,HuGuoGon2023Federated}, they are not directly applicable in asynchronous settings, leading to numerous privacy and security concerns. 
As such, in this section, we focus on gradient reconstruction attacks in AsyncFL, targeting vision models, in which an adversary leverages leaked gradient updates to infer sensitive information about the original training data, reconstructing images or other private inputs with high fidelity.
\subsection{Threat Model and Security Assumptions}
\label{subsec03-01}
We assume an honest-but-curios adversary who could be present either at the server or the client. 
The aggregation server receives asynchronous model updates from clients, who update these model weights using their private training datasets. We assume that the clients' datasets have no overlap, a standard assumption in both SyncFL and AsyncFL literature~\cite{CheNinYue20,Mam21}. The server updates the aggregated model upon receipt of each client's model or on expiration of a countdown timer. 

The adversary is an honest-but-curious party, adhering to the training algorithm without introducing any modifications. The adversary has access to the global model's weights at any time step during the training phase and uses this information and attempts to utilize these weights to reconstruct clients private data. The aggregator employs an existing secure aggregation scheme, preventing the adversary from directly accessing client updates, but we assume that the adversary can access metadata in each training round to uniquely identify the participants participating in the corresponding round of training. 
We note that the adversary does not need to conduct an online attack and can conduct it in a post-training phase by leveraging the recorded global weight updates. This approach avoids any resource utilization or latency anomalies, making the attack stealthier and more difficult to detect.

We use different priors commonly utilized in the literature to analyze \sysname's privacy. These priors range from no prior to an identically distributed distribution to the target distribution.

\subsection{Attack Design} 
\label{subsec03-02}
We consider adversaries with four different types of prior knowledge for our reconstruction attack. The default and weakest attack involves an adversary with {\it \textbf{no prior knowledge (No Prior)}}~\cite{deepLeakage}. The second type consists of adversaries with priors that include the {\it \textbf{total variation (TV)}} of the dataset~\cite{invertGradiants}. The third type involves adversaries with knowledge of the dataset's {\it \textbf{fidelity regularization (Fidelity)}}~\cite{batchRecovery}. Finally, the fourth and strongest type includes adversaries employing a {\it \textbf{deep prior}}, where it has access to an Independent and Identically Distributed (IID) dataset, which it uses to train an autoencoder~\cite{gradientObfuscation}.

Although secure aggregation techniques render access to individual client gradients impractical in synchronous FL, this becomes a practical threat in asynchronous settings.
The increased likelihood of a single client update resulting in a global update enables the adversary to directly exploit the update and extract the individual client’s gradient.
Thus, an adversary with no prior knowledge can leverage the global weights at time step $t$ ($\mathcal{W}^{t}$) and the global weights from the previous time step ($\mathcal{W}^{t-1}$) to compute client $k$'s gradients:
\vspace{-0.15in}
\begin{align*}
\mathcal{W}^{t}  = \mathcal{W}^{t-1} -  \frac{1}{K}*\nabla_{\mathcal{\theta}}^k\mathcal{L}(x^{Priv},y^{Priv})
\\
\nabla_{\mathcal{\theta}}^k\mathcal{L}(x^{Priv},y^{Priv}) = (\mathcal{W}^{t} - \mathcal{W}^{t-1})*{K},
\vspace{-0.05in}
\end{align*}
where $K$ is the total number of clients, and $\nabla_{\mathcal{\theta}}^k\mathcal{L}(x^{Priv},y^{Priv})$ represents the gradient update computed by client $k$ using their private data sample ($x^{Priv}$) and the corresponding private label ($y^{Priv}$). 
Upon successful extraction of the client's local gradients from the global model, the adversary initializes a dummy input and its associated dummy label. By passing these dummy values through the global model update from a previous time step--used by the client to compute its gradient updates--the adversary calculates the gradient of the dummy loss function with respect to the model parameters \( \nabla_{\mathcal{\theta}}^k \mathcal{L}(x^{\text{dummy}}, y^{\text{dummy}}) \). If the adversary possesses no prior about the client, it minimizes the Euclidean distance between the dummy gradients and the client's private gradients by iteratively updating the dummy input and label, by using gradient descent, until they converge to the private input and label. After reconstructing \( x^{\text{Priv}} \), the adversary leverages metadata from the aggregation round, \eg participant ID, to identify the owner of the reconstructed sample.
\begin{figure}[t]
\centering
  \includegraphics[width=0.9\columnwidth]{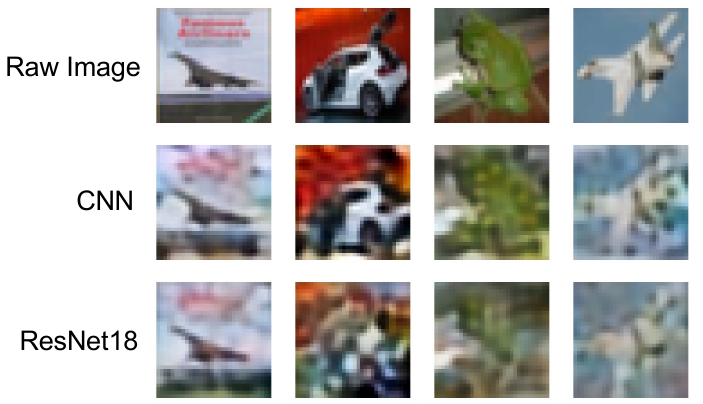} 
  \vspace{-0.15in}
  \caption{Client data reconstruction by an adversary with deep prior knowledge~\cite{gradientObfuscation} across two neural network architectures. These results demonstrate the leakage of sensitive data that takes place in AsyncFed.}
  \label{fig:deep_prior_reconstruction} 
  \vspace{-0.25in}
\end{figure}

For more complex architectures, due to increased complexity and higher parameter count, minimization without any prior becomes infeasible. The second type of adversary (\ie TV), while using a similar setup, uses a cosine similarity loss function, to make optimization less sensitive to local optimas. The adversary utilizes its knowledge of the datasets total variation to add a TV regularization as a loss term. This leads to reduced search space and a higher fidelity reconstruction. 
Stronger adversaries can further utilize their insight into the data distribution to construct stronger regularization, called fidelity regularization, in order to reduce the search space to realistic images. This allows for the elimination of datapoints that have similar variance and parameter updates but lack the fidelity and structure of the target dataset.
\begin{figure*}[t]
\centering
  \includegraphics[width=0.95\textwidth]{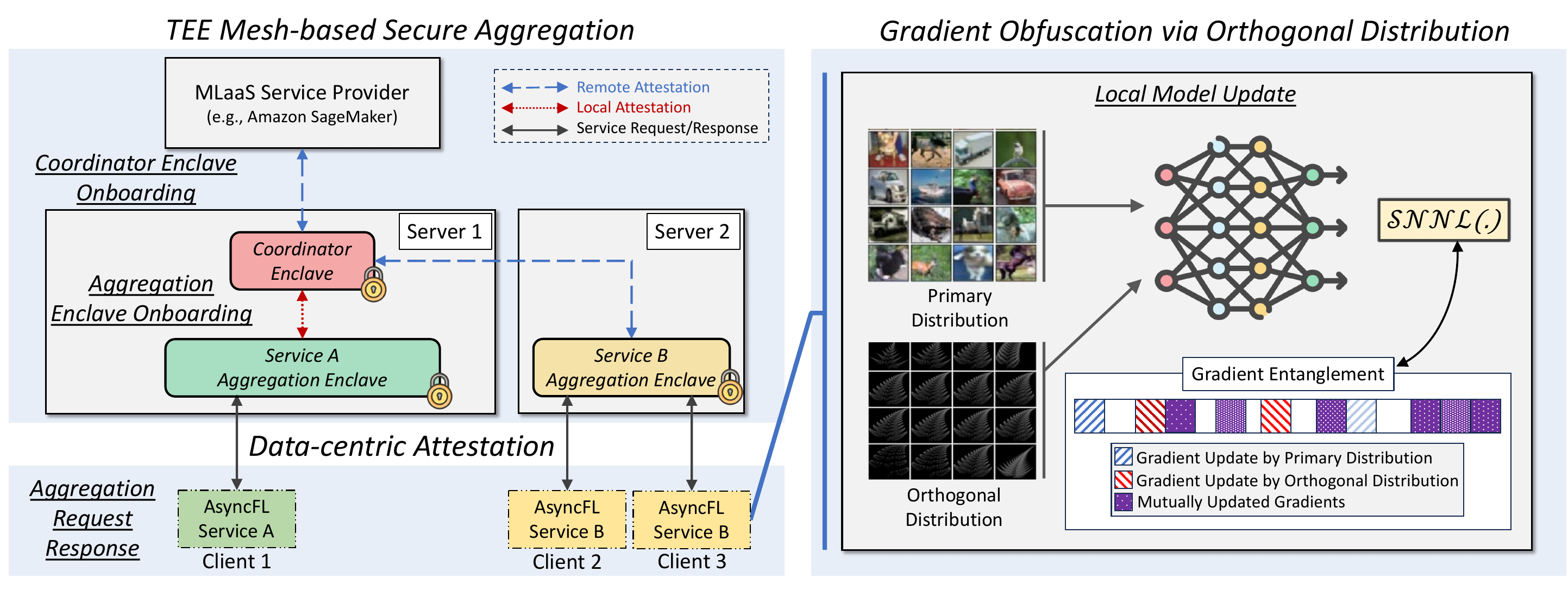} 
  \vspace{-0.1in}
  \caption{The architectural overview of \sysname framework, including its three main building blocks: {\it \textbf{Gradient Obfuscation via Orthogonal Distribution}}, {\it \textbf{TEE Mesh-based Secure Aggregation}}, and {\it \textbf{Data-centric Attestation}}.}
  \label{fig:overview}
  \vspace{-0.1in}
\end{figure*}

Our final adversary, by utilizing higher resources and an IID distribution, trains autoencoders with the goal of reducing the latent space dimensionality. This allows the adversary to perform reconstruction by searching the reduced space and then using the decoded output to calculate the reconstruction loss. While the attack shows a high success rate, we consider this the strongest prior because the adversary needs access to an identically distributed dataset, which may not be available in a federated learning setting. This also makes the deep prior a good worst-case reconstruction test in the FL setting. Figure~\ref{fig:deep_prior_reconstruction} illustrates the outcome of our reconstruction attack, using CNN and ResNet18 architectures with deep prior knowledge, on a set of low-res images, demonstrating attack practicality even in the most unlikely attack scenario. The full attack results will be detailed in Section~\ref{sec08}.

%% file: sec04.tex
\section{Design Overview}
\label{sec04}
We propose a novel AsyncFL framework called \sysname (Figure~\ref{fig:overview}), that integrates three key components: a client-side gradient obfuscation mechanism to safeguard client data, a secure aggregation service mesh for efficient (aggregation) enclave deployment at the edge, and an implicit attestation mechanism for scalable and vendor-agnostic enclave verification. Together, these components satisfy the following design goals (DGs):
    
\noindent $\bullet$ \,{\bf DG 1:} Obfuscate privacy-sensitive information from global model updates while maintaining performance with minimal overhead. 

\noindent $\bullet$ \,{\bf DG 2:} To develop a secure aggregation solution that protects against a malicious server exfiltrating client information, is not bound to a specific hardware vendor, and can utilize TEE hardware.

\noindent $\bullet$ \,{\bf DG 3:} To enable vendor-agnostic enclave attestation, allowing the participating client to authenticate the aggregator irrespective of the TEE provider (\eg Intel SGX or Nvidia Confidential Computing).

\noindent $\bullet$ \,{\bf DG 4:} To develop a more scalable and asynchronous authentication solution than current state-of-the-art frameworks.

\subsection{TEE Mesh-based Secure Aggregation (TMSA)}
\label{subsec04-01}
The first component of the \sysname~framework is {\it TEE mesh-based secure aggregation (TMSA)}, which is a hierarchical enclave structure. The key feature of TMSA is the {\bf abstraction of all hardware-specific attestation primitives to make them vendor independent} and provisioning credentials, and these primitives to a central attestation enclave. 
The first layer of TMSA, the \textit{Coordinator Enclave}, centrally handles all attestation and provisioning required to deploy TMSA's second layer. %
Once the coordinator enclave is deployed and attested by the service provider, it is then given the attestation primitives and can attest and deploy the second layer, the \textit{Aggregation Enclave}.
This second layer, responsible for ML aggregation tasks, spans across multiple servers and utilizes various attestation primitives and methods.

The initial provisioning phase, with the deployment of the coordinator enclave, allows the service provider to utilize cross-platform aggregation enclaves at the aggregation server during the training stage without actively participating in the hardware attestation process. Moreover, running the aggregation process inside an enclave protects client-sensitive gradient updates and metadata from a malicious server. These properties satisfy \textbf{DG 2}. The design of TMSA is detailed in Section~\ref{subsec05-01}.

\subsection{Data-centric Attestation (DCA)}
\label{subsec04-02}
The second component of \sysname~is a novel {\it data-centric attestation mechanism (DCAM)}, which allows clients to attest the aggregation enclaves without relying on any enclave-specific measurements or collaterals using a MABE construct~\cite{ChaCho09}. %
In particular, aggregation enclaves are provisioned with private attributes, \ie credentials needed for secure aggregation, only after the coordinator enclave has successfully attested them. 
This ensures that only trusted enclaves will be provisioned with the correct attributes for aggregation. This enables the key feature of DCAM, which is {\bf allowing the client to perform an implicit attestation of the aggregation enclave} by virtue of the enclave's possession of the correct attributes.
This allows the client to perform the attestation without relying on any hardware-specific or vendor-specific measurement or information. 
The MABE construct will be used in a hybrid encryption scheme to achieve cost-effective secure communication between the client and the aggregation enclave. The client utilizes the attributes provisioned to the aggregation enclave to securely share a symmetric key (\eg AES) with the aggregation enclave for secure data sharing. 
Performing implicit attestation eliminates reliance on enclave-specific measurements or collateral, removing the dependency on specific TEE providers and hardware architectures. Moreover, replacing TLS with MABE-based hybrid encryption speeds up the attestation process, as detailed in Section~\ref{subsec05-02}. These properties satisfy \textbf{DG 3} and \textbf{DG 4}. %

\subsection{Gradient Obfuscation via Orthogonal Distribution (GOOD)} 
\label{subsec04-03}
The final component of \sysname is a novel mechanism, called {\it Gradient Obfuscation via Orthogonal Distribution} (GOOD), designed to protect the privacy of gradient updates. The key feature of GOOD is \textbf{obfuscating the client's private gradient updates by entangling them with the gradients of an orthogonal distribution}--a distribution that shares minimal overlap with the distribution of the primary task. To achieve this, GOOD employs a loss function derived from a modified version of the Soft Nearest Neighbor Loss (SNNL)~\cite{FroPapHin19}. This loss ensures the entanglement between the primary and the orthogonal tasks, making it infeasible to disentangle the gradients of them without any strong priors, enhancing privacy.

The orthogonal distribution may consist of natural images, such as those from the GTSRB dataset~\cite{stallkamp2012man}, or their synthetic variations~\cite{synthImages}. However, in many real-world training scenarios, the availability of these images may be infeasible, or their generation may become prohibitively expensive due to the computational demands of adversarial networks and diffusion models~\cite{ulhaq2022efficient,chen2024opportunities}. To address this, we leverage fractal images generated through Iterated Function Systems (IFS)~\cite{AndFar22}, which offer a scalable and efficient method for dynamically creating the desired orthogonal distribution, either during training or in an offline phase. The low cost of generating fractal distributions enables their real-time generation on client devices with minimal overhead~\cite{chu2000fast,brown2010highly}, thereby satisfying \textbf{DG 1}. The design of GOOD is detailed in Section~\ref{subsec05-03}.

%% file: sec05.tex
\section{Detailed Design and protocols}
\label{sec05}
In this section, we describe the \sysname components, beginning with the gradient obfuscation method (Section~\ref{subsec05-03}), followed by the proposed secure aggregation service mesh structure (Section~\ref{subsec05-01}), and the implicit attestation mechanism (Section~\ref{subsec05-02}). 
We choose this order for better exposition. 
\subsection{Client-Side GOOD Training} 
\label{subsec05-03}
The two key components of GOOD training are fractal distribution generation, which allows gradient obfuscation, and a SNNL regularization loss component, which allows for latent space entanglement of distributions. The fractal distribution is generated by using IFS-based fractal generation algorithms proposed in existing model pretraining literature, where low-cost synthetic images are generated based on natural law. Due to reduced generation cost and no reliance on collected datasets, the client can do a real-time generation of fractal images or can create the distribution for itself in an offline phase~\cite{KatOkaKaz20,AndFar22}.

Introducing the second component (SNNL regularization term) ensures that the orthogonal distribution is effectively entangled with the primary distribution. Without sufficient entanglement, the subspace of the orthogonal distribution in the latent space remains distinct and easily separable from that of the primary task~\cite{JiaChoCha21}. To address this and better intertwine the gradients of the orthogonal task and primary task, we use a variant of SNNL loss to apply a subspace entanglement constraint. This constraint enhances the privacy of the obfuscated gradients by making it infeasible to isolate information specific to the primary task.

Existing entanglement methods using SNNL for learned representations have proven effective in reducing the distance between the representations of a hidden distribution and the primary distribution~\cite{JiaChoCha21}. These methods typically rely on Euclidean distance to bring the manifold representations closer. However, Euclidean distance introduces challenges, as it makes the entanglement measure sensitive to norm magnitudes and can lead to an increased influence of individual datapoints on local optimality, hence not allowing for maximal possible overlap. Similar issues make Cosine Similarity a better alternative for clustering problems~\cite{Hua08}. We modify the SNNL loss by using cosine similarity as the distance measure between latent space representations instead of Euclidean distance, leading to better alignment due to feature normalization~\cite{ChuJiaLei17} and increased overlap between the subspaces in the latent space.

\floatname{algorithm}{Algorithm}
\setcounter{algorithm}{0}
\begin{algorithm}[t]
\small
\caption{GOOD-based Client Training }
\label{Algo-1}
\textbf{Input:} \raggedright ${W}_k$,  $\mathbb{D}_C^{Priv}$, $\tau \in [0,1]$, ${N, b}$ (Fractal Parameters).\\
\textbf{Output:} ${W}_k$
\begin{algorithmic}[1]
\vspace{-0.1in}
\item[]
\Statex \hspace{-0.2in} \textcolor{SeaGreen}{\(\triangleright\) Generate Fractal Distribution}
\STATE ${D}_C^O$ = IFS\_GENERATOR(${N, b}$) 

\Statex \hspace{-0.2in} \textcolor{SeaGreen}{\(\triangleright\) Start Training}

\FOR{$i = 1$ \textbf{to} $n_{\text{epochs}}$}

\Statex \hspace{-0.2in} \textcolor{SeaGreen}{\(\triangleright\)Sample Primary and Orthogonal Distributions}

\STATE $\{X_T, Y_T\} = \{X_C^{Priv} \cup 
\vspace{0.05in}
X_C^O,Y_C^{Priv} \cup Y_C^O\}$ \newline
$,\forall \{X_C^{Priv},Y_C^{Priv}\}  \in \mathbb{D}_C^{Priv}, \forall \{X_C^O,Y_C^O\}  \in \mathbb{D}_C^O$

\Statex \hspace{-0.2in} \textcolor{SeaGreen}{\(\triangleright\)Entanglement-based Loss Computation}

\STATE $\mathcal{L} = \mathcal{L_{CE}}\big({F}_S(X_C^{Priv}),Y_C^{Priv} \big)$ + $\tau \, C$-$SNNL\big({F}_S(X_T),Y_T \big)$

\STATE $W_k^{(t+1)} = W_k^{(t)} - \eta \nabla_{W_k} \mathcal{L}(W_k^{(t)}, X_T, Y_T)
$

\ENDFOR
\STATE {\bf Return $W_k$ } 
\end{algorithmic}
\end{algorithm}

We further modify the SNNL function by introducing a lower bound of zero and replacing the exponential scaling with a logarithmic scale, imposing a greater penalty for lower spatial overlap compared to SNNL. The final regularization term looks as follows:
\begin{align*}
\mathcal{C-SNNL} &  = -\sum\limits_{i \in 1...b } \log\Biggl( \frac{\sum\limits_{\substack{j \in 1..b \\ j \neq i \, \land \, y_{i} = y_{j}}} \log {\Big(\mathcal{C.S}\big(x_i, x_j\big)\Big)}}{\sum\limits_{\substack{k \in 1..b \, \land \, k \neq i }} \log {\Big(\mathcal{C.S}\big(x_i, x_k\big)\Big)}}\Biggl),
\end{align*}
where $\mathcal{C.S}()$ is the cosine similarity measure.
In summary, the proposed GOOD mechanism combines fractal generation using IFS and latent space entanglement to achieve gradient privacy while maintaining model performance. Fractal images offer a scalable and cost-effective means of generating an orthogonal distribution, while latent space entanglement effectively obfuscates client gradients, helping preserve model performance and also enhancing its discriminative capabilities~\cite{FroPapHin19}.

\begin{figure*}[!ht]
\centering
  \includegraphics[width=0.9\textwidth]{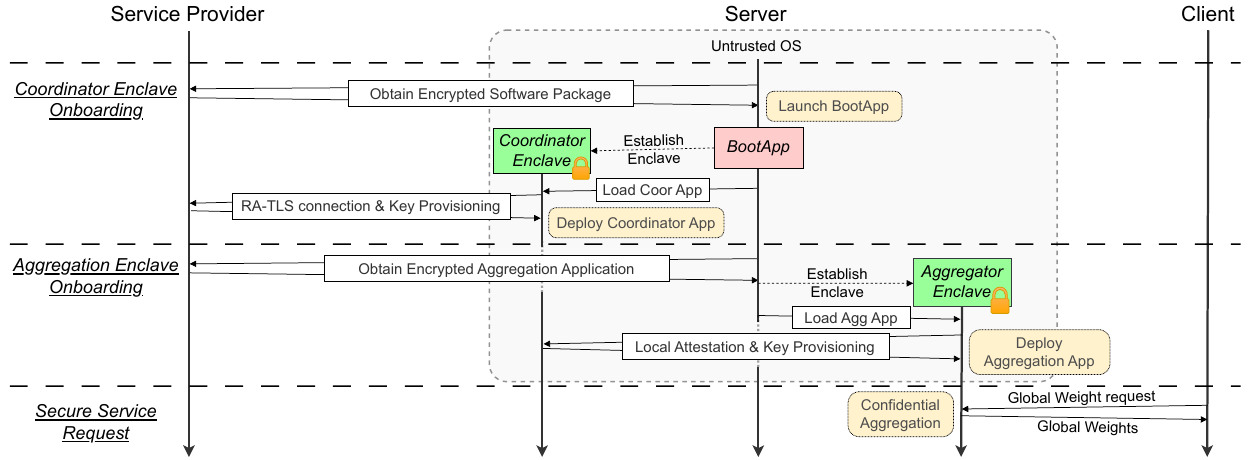} 
  \vspace{-0.1in}
  \caption{The sequence diagram of \sysname outlines the communication flow between the Service Provider, Server, and Client. Detailed interactions and credential establishment processes are presented in Subsections~\ref{subsec05-01} and~\ref{subsec05-02} (Protocols 1–5).}
  \label{fig:DcTC} 
  \vspace{-0.0in}
\end{figure*}

\subsection{TEE Mesh-based Secure Aggregation}
\label{subsec05-01}
Figure~\ref{fig:DcTC} illustrates the interactions between entities in \sysname for onboarding the coordinator and aggregator enclaves.
For simplicity, we illustrate the coordinator enclave and aggregator enclave running on the same edge server. However, \sysname's protocols allow for the running of coordinator and aggregation enclaves on individual servers, with multiple aggregation enclaves onboarded by a coordinator enclave with the right attributes for MABE.
\bigskip

\noindent
{\bf System Setup and Service Registration (Protocol~\ref{proto1})} 

\noindent
A service provider creates the ABE public and private keys ($M_{pk}$, $M_{sk}$) corresponding to the $n$ attributes it wants to support. The service provider then creates a manifest, known as the application's app structure ($AS$), outlining resource requirements and enclave environment specifications for the coordinator and aggregation applications. A token and signature are also included with the manifest to prove authenticity. The $AS$ for the coordinator and aggregation applications are referred to as the coordinator app structure ($CAS$) and the aggregation app structure ($AAS$), respectively (Lines 1-3).

Along with the $AS$, the service provider will also generate the attestation primitives ($attestPrim$) for each application. These attestation primitives will be used during RA-TLS connections to ensure that the applications in the secure enclaves are set up correctly. A set of signature verification and generation keys ($VK_{SP}$, $SK_{SP}$), are generated for communication purposes. The executables for the coordinator and aggregation applications ($CoorApp$ and $AggApp$) are encrypted using symmetric encryption keys $K_{C}$ and  $K_{A}$ respectively to get $encCoorApp$ and $encAggApp$. Encrypting these applications provides confidentiality in case they are distributed ahead of time and enables in-network caching (Lines 4-6). 
\bigskip
\floatname{algorithm}{Protocol}
\setcounter{algorithm}{0}
\begin{algorithm}[t]
\small
\caption{Service Setup}
\label{proto1}
\begin{algorithmic}[1]
\item[]
\begin{center}
\end{center}
\STATE $\Big(\mathsf{sysparam}$$,M_{pk} = (apk_{1},$$\dots, $$apk_{n}),$$  M_{sk} = $$(ask_{1},$$ \dots,$$ ask_{n}\Big)$$ \leftarrow \setup(1^{\lambda}, n)$
\STATE $CAS = \{manifest_{coor}, token_{coor}, signature_{coor}\} $

\STATE $AAS = (manifest_{agg}, token_{agg}, signature_{agg})$

\STATE $encCoorApp  = Enc_{K_C}(CoorApp)$
\STATE $encAggApp  = Enc_{K_A}(AggApp)$

\STATE $\mathsf{Generate}$ $attestPrim$ (Used during RA-TLS) 
\end{algorithmic}
\end{algorithm}

\noindent
{\bf Coordinator Onboarding (Protocol~\ref{proto-2})}

\begin{algorithm}[t]
\small
\caption{Coordinator Enclave Onboarding}
\label{proto-2}
\begin{algorithmic}[1]
\item[]

\begin{center}
\graybox{
\COMMENT {\textbf {At Service Provider}}
}
\end{center}
\STATE $Res \leftarrow   \{BootApp, encCoorApp,  CAS\}$, $\sigma_{Res}\leftarrow Sign_{SK_{sp}}(Res)$.
\STATE $\mathsf{Send}$ ($Res,\sigma_{Res}$) to Server
\begin{center}
\graybox{
\COMMENT {\textbf {At Server}}
}
\end{center}
\STATE $\mathsf{Receive}$  $(Res, \sigma_{Res})$
\STATE $(BootApp, encCoorApp, CAS) \gets$ Extract $Res$.
\IF {$true \gets \mathsf{Verify}_{VK_{sp}}(Res, \sigma_{Res})$}
\STATE $\mathsf{Execute}$ $BootApp$, 
\STATE $CoorEnc$ $\gets$ $BootApp$ initializes a secure enclave for the Coordinator.

\STATE $BootApp$ loads $encCoorApp$ and $CAS$ onto $CoorEnc$.
\ELSE
    \STATE Drop connection and return error 
\ENDIF
\begin{center}
\graybox{
\COMMENT {\textbf {At Coordinator Encalve}}
}
\end{center}
\STATE $\mathsf{Initialize}$ RA-TLS connection with $ServiceProvider$
\begin{center}
\graybox{
\COMMENT {\textbf {At Service Provider}}
}
\end{center}
\IF {$true \gets$  RA-TLS.Authentication()}
\STATE $\mathsf{Return}$ $K_{C}$ to $CoorEnc$
\ELSE
    \STATE Drop connection and return error 
\ENDIF

\begin{center}
\graybox{
\COMMENT {\textbf {At Coordinator Encalve}}
}
\end{center}

\STATE $CoorApp = {Dec}_{K_C}(encCoorApp)$.
\STATE $\mathsf{Execute}$ $CoorApp$
\STATE $\mathsf{Initialize}$ RA-TLS connection with $ServiceProvider$

\begin{center}
\graybox{
\COMMENT {\textbf {At Service Provider}}
}
\end{center}
\IF {$true \gets$  RA-TLS.Authentication()}
\STATE $[A_e] \gets ABE.KeyGen(M_{sk}, e)$
\STATE  $\mathsf{Store}$ $\{e, [A_e ]\}$ in $serverTable$
\STATE $\mathsf{Return}$ $[[A_e], K_A] $ to $CoorEnc$
\ELSE
    \STATE Drop connection and return error 
\ENDIF

\end{algorithmic}
\end{algorithm}

\noindent
Edge servers register with the service provider to provision coordinator enclave by receiving a prepared software package ($Res$). $Res$ contains the $BootApp$, the $encCoorApp$, and $CAS$ (Lines 1-4). %
Upon verifying the integrity of $Res$, the server will execute the $BootApp$ which will then create a secure enclave ($CoorEnc$) for the coordinator, onto which $encCoorApp$ will be loaded. The enclave will then use a high-level RA-TLS connection with the service provider for remote attestation, after which it will be provided with a decryption key ($K_C$) for the coordinator app (Lines 5-18). 

After being initialized, $CoorApp$ then creates a mid-level RA-TLS connection to the service provider to receive the ABE decryption keys. We note that the reason for this second mid-level RA-TLS connection is to verify the legitimacy of the coordinator enclave by the service provider, hence avoiding an enclave-swapping attack. In enclave swapping attack, an attacker can start with the legitimate coordinator enclave and swap it with a malicious coordinator enclave, hence allowing it to steal decryption keys. %
Upon verification of the running coordinator enclave, the service provider proceeds with provisioning the attestation collateral, the decryption key for the aggregation application ($K_A$), and the appropriate private attributes ($[A_e]$). The attributes provisioned by the service provider are generated specifically for use by the coordinator and its cluster and allow for the decryption of client data. The service provider creates a mapping between the cluster ID ($e$) and $[A_e]$ (Lines 19-27).

Once $CoorApp$ is fully provisioned, it will notify the $BootApp$ that it is ready to onboard aggregator enclaves.
The coordinator will be responsible for remote and local attestation and for key provisioning of aggregator enclaves. Local attestation is performed when the coordinator and aggregator enclaves are on the same machine, while remote attestation is performed when they are on separate machines. The process for local and remote attestation is outlined in~\cite{Anati2013InnovativeTF}. With local attestation, the aggregator will generate an enclave report via a hardware instruction that can be directly sent to the coordinator for verification. In the case of remote attestation, the aggregator enclave generates an enclave report, which will be sent to the {\it Quoting Enclave}, a special enclave enabled by Intel SGX. The Quoting Enclave will then locally attest the aggregator enclave and return an SGX quote to the aggregator. This quote is then sent to the coordinator who will use the $attestPrim$ function to verify successful aggregator deployment.
\bigskip

\begin{algorithm}[t]
\small
\caption{Aggregation Enclave Onbording}
\label{proto-3}
\begin{algorithmic}[1]
\item[]

\begin{center}
\graybox{
\COMMENT {\textbf {At Service Provider}}
}
\end{center}

\STATE $Res \gets \{BootApp,EncAggApp, AAS\}$, $\sigma_{Res}\gets Sign_{SK_{sp}}(Res)$
\STATE $\mathsf{Return}$ $\{Res,\sigma_{Res}\}$

\begin{center}
\graybox{
\COMMENT {\textbf {At Server}}
}
\end{center}

\STATE $\mathsf{Receive}$  $\{Res,\sigma_{Res}\}$
\STATE  $\{BootApp,EncAggApp, AAS\} \gets Extract Res$.
\IF {$true \gets \mathsf{Verify}_{VK_{sp}}(Res, \sigma_{Res})$}
    \STATE $\mathsf{Execute}$ $BootApp$, 
    \STATE $AggEnc$ $\leftarrow$ $BootApp$ initializes a secure enclave for the service.
    \STATE $BootApp$ loads $encAggApp$ and $AAS$ onto $AggEnc$.
    \STATE $BootApp$ sends a request to $CoorEnc$ to attest $AggEnc$
\ELSE
    \STATE Drop connection and return error 
\ENDIF
\begin{center}
\graybox{
\COMMENT {\textbf {At Coordinator Enclave }}
}
\end{center}
\IF{attestation successful}
    \STATE $\mathsf{Send}$ $[[A_e], {K_A}]$ to $AggEnc$
\ELSE
    \STATE Drop connection and return error 
\ENDIF

\begin{center}
\graybox{
\COMMENT {\textbf {At Aggregation Enclave}}
}
\end{center}

\STATE $AggApp \gets {Dec}_{K_A}(EncAggApp)$

\State $\mathsf{Execute}$ $AggApp$

\end{algorithmic}
\end{algorithm}

\noindent
{\bf Onboarding Aggregator Enclave (Protocol~\ref{proto-3})}

\noindent
A server begins the onboarding process of the aggregation service by sending a request to the service provider. If the device is not collocated with the coordinator, it will also request the $BootApp$. The response to this request will include $EncAggApp$, the $AAS$, and a signature for verification (Lines 1-4). If not already running, the server initializes the $BootApp$, which in turn, initializes a secure enclave ($AggEnc$). The $BootApp$ then loads the $EncAggApp$ and $AAS$ to the enclave. The $BootApp$ then sends a request to the $CoorEnc$ to attest the newly created enclave (Lines 5-12).
The coordinator then attests $AggEnc$ using local or remote attestation, based on where the enclave is hosted. An inter-enclave secure channel is then established to provision the required keys and attributes (Lines 13-17), allowing $AggEnc$ to decrypt and deploy $AggApp$. 
\bigskip

\noindent
{\bf User Registration} %

\noindent
A client can utilize the aggregation services provided by the service provider once it retrieves the appropriate public attributes and $M_{pk}$. Depending on the service provider these attributes may be freely accessible, or the client can gain access to them after going through a registration process with the service provider. After which the client can securely receive a tuple that includes the public attributes, $M_{pk}$, and the signature of the service provider.

\subsection{Data-centric Attestation} 
\label{subsec05-02}
{\bf Aggregation Request (Protocol~\ref{proto-5})}

\noindent
In the service offloading phase, $AggApp$ is loaded with sufficient credentials and is ready to serve client requests.
Client $u$ is a consumer of the aggregation service running at $AggApp$. We assume that $u$ has already registered for the aggregation service with the provider and obtained sufficient credentials, keys, and policies needed for data encryption. Client $u$'s initial request to $AggApp$ will be to attain a copy of the global weights. In subsequent requests, the client will include its local weights so that they may receive the result of their integration into the global model stored at the aggregator. Client $u$ then encrypts its request which includes its private data (i.e., local weights) using a hybrid cryptosystem. More specifically, $u$ first encrypts its Data using a symmetric key cryptosystem such as AES 256 with key $K_U$ to generate $C_1$. Then $C_2$ will be generated by encrypting $K_U$ using ABE with a public key $M_{pk}$ and a user-defined access policy comprising a subset of the public attributes represented by $\Upsilon_{Agg}$ in a disjunctive boolean clause to control access to the ciphertext (\eg (Aggregation $\wedge$ Microsoft) (Lines 1-8).

\begin{algorithm}[!t]
\small
\caption{Client Aggregation Request}
\label{proto-5}
\begin{algorithmic}[1]
\IF {$W_k == \phi$}
\STATE $U \gets$  Request for global weights
\ELSE
\STATE  $W_k^{Pre} \gets W_k$
\STATE $U \gets \{W_k^{Pre}, W_k\}$
\ENDIF
\STATE $C_1 \gets \mathsf{Enc}_{K_U}(U)$
\STATE $C_2 \gets \encrypt(M_{pk}, \Upsilon_{Agg} , K)$
\STATE $\mathsf{Set}$ $Req \gets \{C_1, C_2, Cert_u\}, \sigma_{Req} \gets \mathsf{Sign}_{SK_u}(Req)$ 
\STATE $\mathsf{Send}$ $\{Req, \sigma_{Req}\}$ to {\it AggEnc} 
\end{algorithmic}
\end{algorithm}

Thus, only allowing the entities with the corresponding attribute keys to decrypt the ciphertext. Client $u$ forms a request made up of $C_1$, $C_2$, and a certificate ($Cert_u$) for verification purposes, which is then signed using $SK_u$ before sending to $AggEnc$ (Lines 9-10). 
\bigskip

\noindent
{\bf Aggregation Response (Protocol~\ref{proto-6})}
\\
\noindent
When $AggApp$ receives $u$'s request, it first verifies the client's signature for correctness (Lines 1-3). $AggApp$ then decrypts $C_2$ using its private attributes, $[A_{e}]$ (using CP-ABE), thereby retrieving the symmetric key $K_U$ which is then used to decrypt $C_1$. The aggregator now has $u$'s Data and can satisfy the request. If the request is not valid then the connection is dropped. Otherwise, based on the request received $AggApp$ either sends a copy of its global weights to the user or incorporates the client's weights into its model before sending a copy of the result to $u$ (Lines 4-10). The result is encrypted using $K_U$ (i.e., $C_3$) and then sent back to $u$ along with the certificate of the provisioning enclave (Lines 11-12). After receiving the response, $u$ decrypts $C_3$ and obtains the global weights.

For further security, an enclave report for $AggEnc$ may be obtained which can be verified by $u$ using the service provider's certificate, ensuring that the aggregation was executed correctly and from inside of a secure enclave. The provisioning enclave certificate can be validated using Intel’s root of trust which is already cached at the client. 
We note that checking the enclave report or verifying the chain of trust is not necessary. By receiving the computation result, encrypted with the client’s selected symmetric key ($K_U$), we can infer that a legitimate enclave with requisite credentials (ABE keys) had decrypted the symmetric key and the client’s data.

\begin{algorithm}[t]
\small
\caption{Aggregation Server's Response to Client}
\label{proto-6}
\begin{algorithmic}[1]
    
\item[]
\begin{center}
\end{center}

\STATE $\mathsf{Receive}$  $\{Req, \sigma_{Req}\}$
\STATE $\mathsf{Extract}$ $\{C_1, C_2, Cert_u\} \leftarrow Req$, $VK_{u} \leftarrow Cert_{u}$
\IF {$true \leftarrow \mathsf{Verify}_{VK_{u}}(Req, \sigma_{Req})$}

\STATE{$ K \leftarrow \decrypt(M_{pk},[A_e],C_2)$}
\STATE $U = \mathsf{Dec}_{K_U}(C_1)$
    \IF{$U$ is a request for global weights}

    \STATE $C_3 =  \mathsf{Enc}_{K_U}(W_G)$, 
    
    \STATE $\mathsf{Return}$ $C_3$  
    \ELSIF{$U$ is a aggregation request}%
        \STATE $W_G  = W_G -  \frac{1}{K}*(W_K^{Pre} - W_K)$
        \STATE $C_3 =  \mathsf{Enc}_{K_U}(W_G)$, 
    
        \STATE $\mathsf{Return}$ $C_3$          
    \ENDIF

\ELSE
    \STATE Drop connection and return error 
\ENDIF

\end{algorithmic}
\end{algorithm}

%% file: sec07.tex
\begin{figure*}[h]
\centering
  \includegraphics[width=\textwidth]{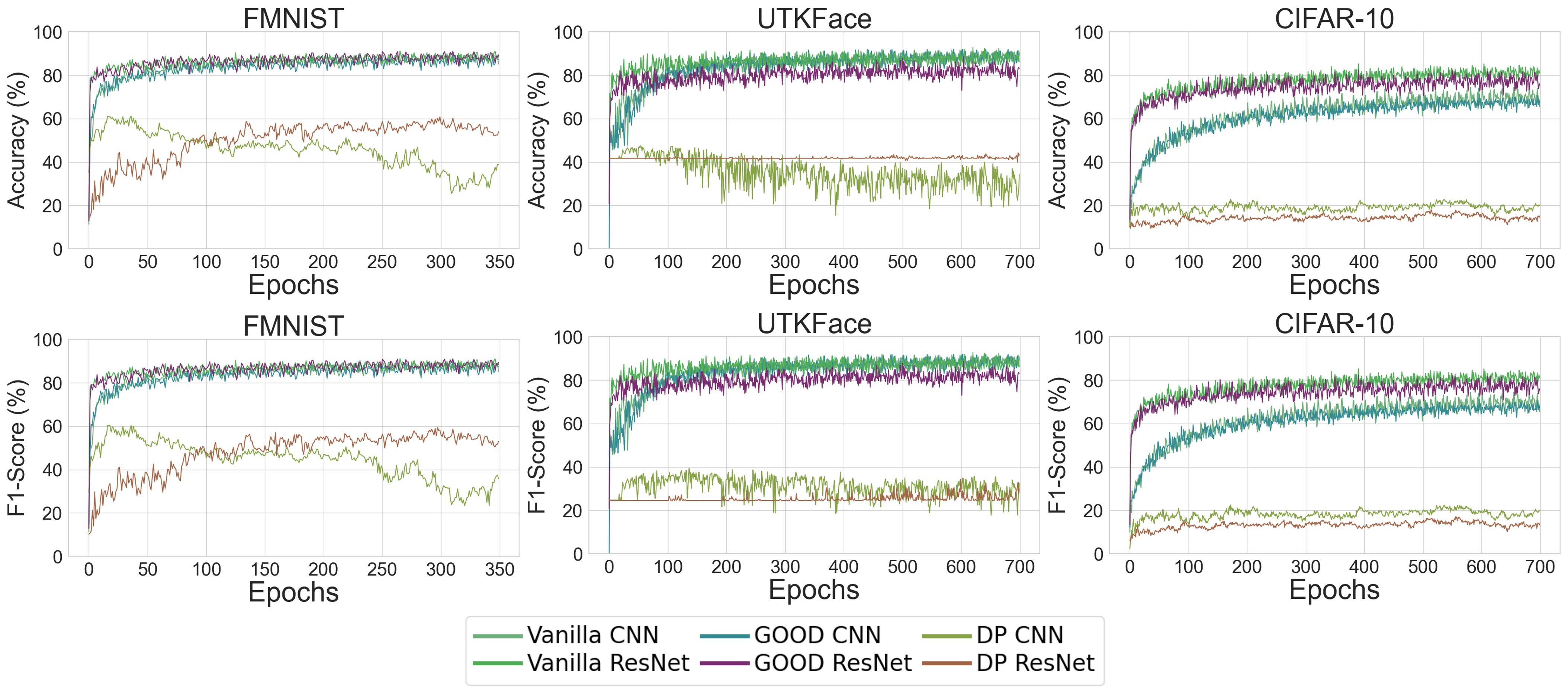} 
  \vspace{-0.23in}
  \caption{Vanilla vs GOOD based asynchronous federated learning.}
  \label{fig:detection_correction} 
  \vspace{-0.15in}
\end{figure*}

\section{SECURITY ANALYSIS}
\label{formalsec} 
We prove the security of our constructions in the well-known Universal Composability (UC) framework~\cite{Canetti01}. The UC paradigm captures the conditions under which a given distributed protocol is secure, by comparing it to an ideal realization of the protocol. To accomplish this, the UC framework defines two “worlds”, firstly, a real-world, where the protocol $\pi$, to be proved secure, runs in the presence of a real-world adversary, $\adv$. Secondly, the ideal world, where the entire protocol $\phi$ is executed by an ideal and trusted functionality, in the presence of a simulator, $\simu$, which models the ideal-world adversary. All users only talk to an ideal functionality via secure and authenticated channels, the ideal functionality takes input from users, performs computations and other operations, and returns the output to the calling users. Our goal is to prove that no entity represented by $\env$, can successfully distinguish between the execution of the two worlds.

\subsection{Design of Ideal Functionalities}
We define an ideal functionality, $\fdctc$, consisting of five independent ideal functionalities, $\fregister, \fresp, \ftee, %
\fsmt, \fsig$. %
Specifically, $\fregister$ models the service setup, coordinator enclave and aggregate enclave onboarding processes, as well as encryption and decryption functionality for symmetric key encryption and attribute-based encryption. $\fresp$ models the user's request for the aggregation service and the response of the edge server to the user. We use the helper functionality $\ftee$~\cite{pass2017formal} to model the ideal functionality of a secure enclave.%
We also use two other helper functionalities, $\fsig$~\cite{canetti2004universally} and $\fsmt$~\cite{Canetti01}, to model ideal functionalities for digital signatures and secure/authenticated channels, respectively.
We assume that $\fdctc$ maintains an internal state that is accessible at any time to all the ideal functionalities.
We describe the functionalities of $\fdctc$, discuss some of their motivating design choices, and give the proof of the following theorem in Appendix~\ref{sec:full-sec-analysis}.

\begin{theorem}
\label{thm:uc}
 Let $\fdctc$ %
 be an ideal functionality for \sysname. Let $\adv$ be a probabilistic polynomial-time (PPT) adversary for \sysname, and let $\simu$ be an ideal-world PPT simulator for $\fdctc$. %
 \sysname UC-realizes $\fdctc$ for any PPT distinguishing environment $\env$.
\end{theorem}

%% file: sec08.tex
\section{Experimental Results and Analysis}
\label{sec08} 
In this section, we first elaborate on the \sysname' implementation scope, and review the experimental setup. We will then share the results and findings, including the proposed attack, mitigation strategy, and the system performance and scalability of the proposed data-centric attestation.

\subsection{System Implementation Scope}
\label{implementation}
To evaluate our approach, we implemented the \sysname~framework, consisting of an aggregation application designed to run on a server and a client application intended for operation on client devices. For comparison, we also implemented an alternative framework employing the same set of applications but utilizing the standard RA-TLS attestation framework~\cite{intel2018ratls}.
Our \sysname~implementation adheres to Protocols~\ref{proto-5} and~\ref{proto-6}. We assumed that the necessary credentials had already been distributed to both the client and the server. Hence, we partially implemented Protocols~\ref{proto1} through~\ref{proto-3}. This does not have an impact on our performance comparison because, in practice, server and client onboarding operations occur at much lower frequencies than client service requests. We also implemented GOOD-based client training (Algorithm~\ref{Algo-1}) and compared it against vanilla and differential privacy-based AsyncFL approaches.

All implementations were written in Python and executed using Gramine~\cite{gramine}, a library OS that allows unmodified applications to run within an SGX enclave. The SGX enclaves were configured with 4 GB of memory and 32 threads. We extended the C++ MABE implementation from~\cite{edgeCom}, generalizing it to accept variables specifying the number of authorities and number of attributes from each authority in the system. Additionally, \sysname leverages the reliable UDP implementation of MsQuic, a C++-based QUIC protocol implementation, to facilitate communication. Both aggregator implementations include federated learning mechanisms capable of integrating client updates.
In both implementations, the aggregator hosts the global weights, which are distributed to clients upon request. The interaction begins when a new user (client) requests the global weights from the aggregator. The aggregator responds by providing an ID and the current global weights. The client then updates the weights using its local data and sends the updated weights along with its ID back to the aggregator. Thereafter, the clients in the system can request the newly updated global weights.

\subsection{Dataset and Architecture}
We evaluate the reconstruction attack and the efficacy of the proposed gradient obfuscation defense using these datasets:

$\bullet$ {FMNIST}~\cite{FMNIST}, a dataset consisting of $70,000$ grayscale images. These images are labeled under ten clothing categories such as sneaker, shirt, pullover, etc.

$\bullet$ {CIFAR-10}~\cite{CIFAR}, a dataset consisting of 60,000 color images labeled under ten classes such as airplane, automobile, bird, etc. There are $50000$ training and $10000$ test images.

$\bullet$ {UTKFace}~\cite{UTKFace}, a multi-task dataset consisting of $20,000$ face images. Each image is labeled based with age, gender, and ethnicity. We train our classifiers to classify the images based on ethnicity.

We utilize two architectures to train our target models, namely, a SimpleCNN architecture (consisting of three convolution layers and two dense layers) and a Resnet18~\cite{resnet18} architecture. Both of these architectures are trained on all of the aforementioned datasets.
{\color{black} For the CIFAR-10 and UTKFace datasets, we set the number of epochs to 700; for the FMNIST dataset, we set EPOCHS to 350. We set the training batch size for all models to 64, used RMSprop and Adam for the optimizers, and used CrossEntropyLoss for the loss function for vanilla training.}
We compare GOOD against differential privacy (DP-SGD). We train the differentially private models on an epsilon value 0.1 with gradient norm of 1.2 and delta of $10^{-5}$. Lower epsilon values improve the level of privacy in a model but also adversely impact the utility of the said model. In the following sections we analyze both the utility performance and privacy performance of GOOD and DP-SGD. 

\begin{figure}[t]
\centering
  \includegraphics[width=\columnwidth]{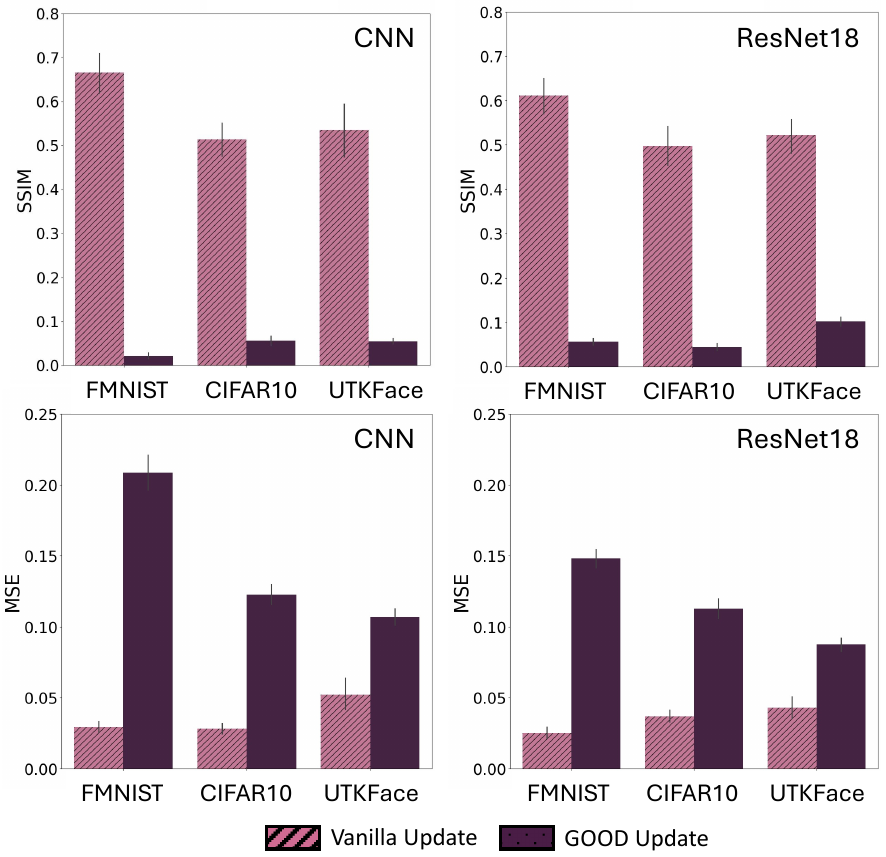} 
  \vspace{-0.24in}
  \caption{Comparison of the GOOD and vanilla approaches against a data reconstruction attack using the strongest prior knowledge. The results highlight GOOD's superiority in effective attack mitigation, even under the strongest prior.}
  \label{fig:Reconstruction_deep_prior} 
  \vspace{-0.15in}
\end{figure}

\begin{figure*}[htb]
    \centering %
\subfloat[ GOOD reconstruction.\label{fig:GOODRecon}]{
  \includegraphics[width=0.48\textwidth]{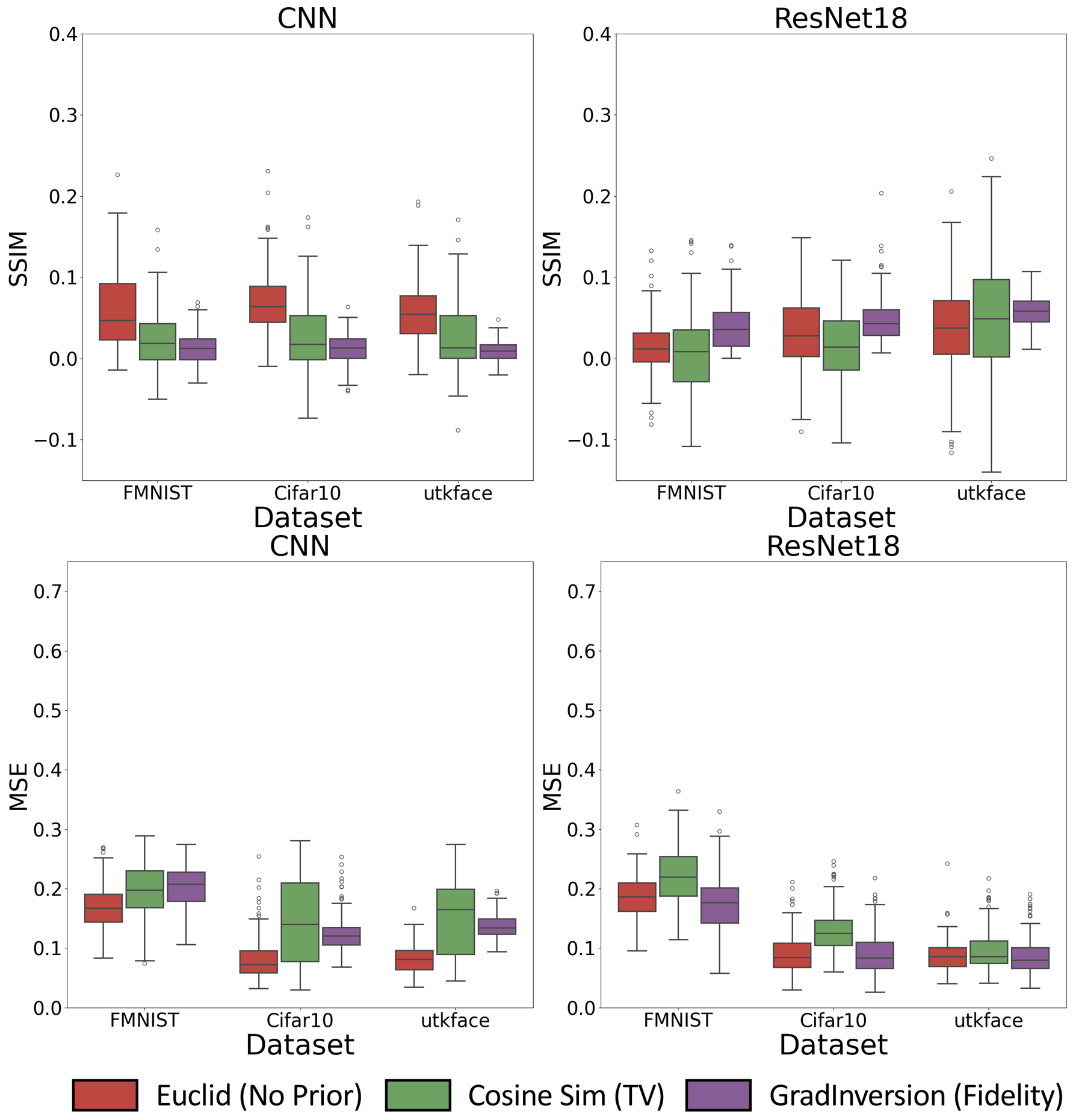}
}
\hspace{-0.5in}
\hfill
\subfloat[DP-SGD reconstruction. \label{fig:DPRecon}]{
  \includegraphics[width=0.48\textwidth]{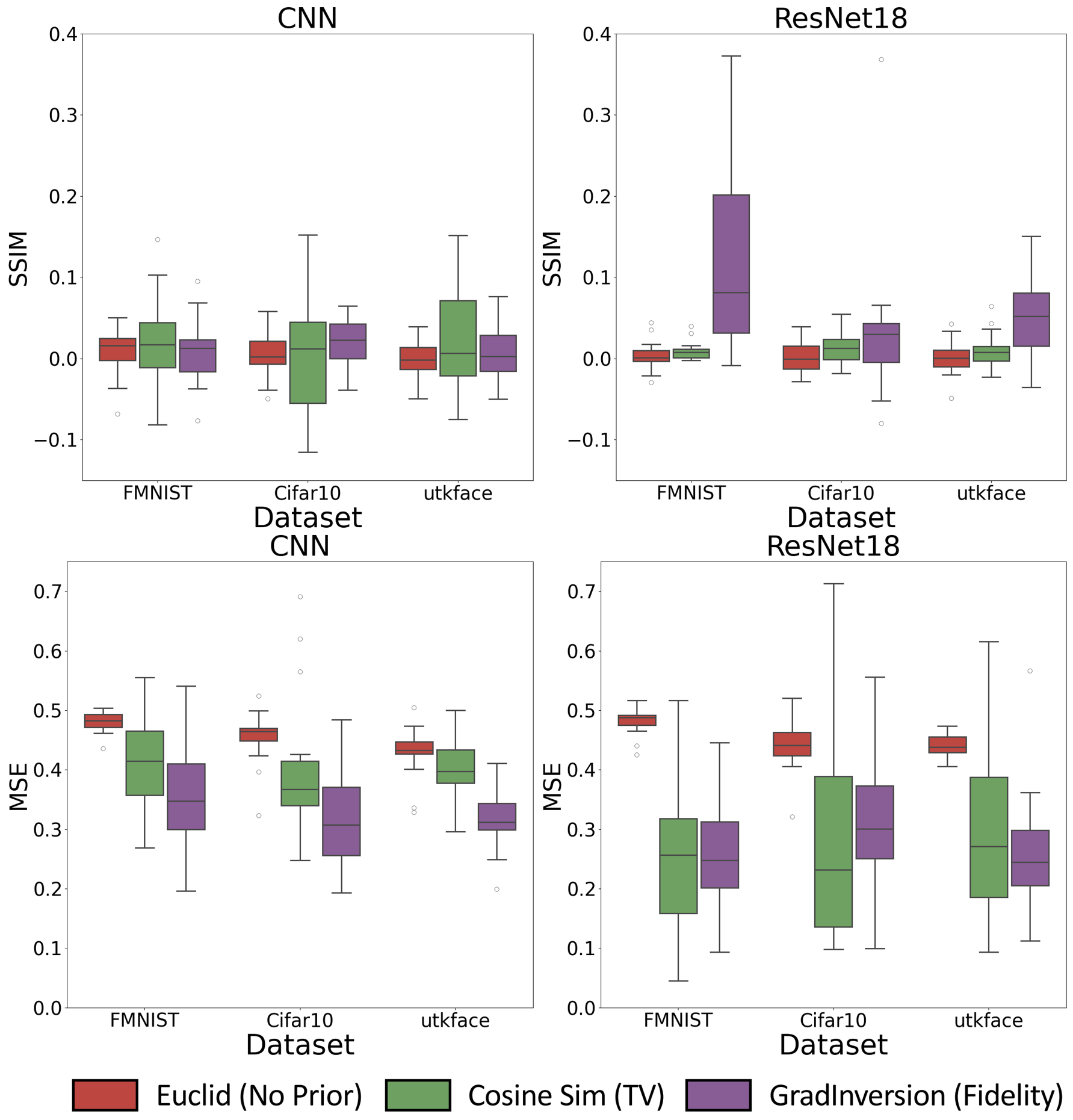}
}
\vspace{-0.15in}
\caption{Effectiveness of GOOD in mitigating data reconstruction attacks under varying levels of prior knowledge. On average, GOOD maintains consistently low SSIM values across all scenarios, highlighting its robust privacy-preserving capabilities.}
\label{fig:Reconstruction_different_prior}
\vspace{-0.1in}
\end{figure*}

\subsection{Utility Performance}
We first assess the utility of GOOD when compared to the other schemes. Figure~\ref{fig:detection_correction} illustrates the accuracy and F1-score of GOOD and vanilla asynchronous federated learning across the dataset and architecture combinations. The results are plotted for every ten epochs. 
From the figure, it is evident that the training performance using the GOOD-based approach aligns closely with the vanilla approach in terms of both accuracy and convergence rate. The only exception is observed with the UTKFace dataset, where the overall accuracy and F1-score are slightly lower compared to the vanilla approach, despite a similar convergence rate. We attribute this discrepancy is to the inherent characteristics of the dataset. 
We observed that the GOOD-CNN architecture performs slightly worse than Good-ResNet18, requiring more time to converge to a marginally lower value. This indicates that a more advanced architecture, like ResNet18, may be preferable for achieving optimal results in both approaches. Overall, our findings show that our gradient obfuscation approach maintains model utility without compromise. Next, we discuss the privacy benefits of GOOD in mitigating data reconstruction attacks.
Figure~\ref{fig:detection_correction} also shows the training performance of the DP-SGD approach. We observe significantly lower accuracy and F1 score with DP-SGD compared to both the GOOD and vanilla approaches. We see that FMNIST and UTKFace perform better than Cifar10 when trained using DP-SGD. This is mainly due to the higher complexity of the Cifar10 dataset.

\subsection{Privacy-preserving Performance}
We also evaluated the privacy benefits of our proposed gradient obfuscation method by comparing it to the vanilla asynchronous FL approach using structural similarity index (SSIM) and mean square error (MSE) metrics. Figure~\ref{fig:Reconstruction_deep_prior} illustrates the performance comparison of GOOD and the vanilla approach against an adversary equipped with the strongest prior knowledge~\cite{gradientObfuscation}, \ie access to non-overlapping IID shadow dataset. Specifically, it can be observed that the vanilla approach results in significantly higher SSIM values compared to GOOD--approximately \textbf{800\% higher}--across all dataset and architecture combinations. The MSE scores for GOOD are observed to be \textbf{nearly 300\% higher} than those of the vanilla approach across all datasets, except for UTKFace, where the MSE scores are still \textbf{around 200\% higher}. These results indicate that reconstruction attempts under GOOD produce significantly less perceptible outcomes, effectively safeguarding client privacy even when the attacker possesses strong prior information.
We also observed that while Cifar10 and UTKFace show similar SSIMs and MSE scores for the GOOD update, FMNIST shows a noticeably higher MSE for both CNN and ResNet18, suggesting that simpler datapoints are easier to obfuscate, resulting in a greater reduction in reconstruction effectiveness. However, when left unobfuscated, simpler data is the easiest to reconstruct. These findings demonstrate that our approach effectively protects the privacy of the client regardless of the complexity of the data.

We also evaluated the effectiveness of data reconstruction attacks leveraging various prior knowledge, as outlined in Section~\ref{subsec03-02}, against the GOOD method and DP-SGD (Figure~\ref{fig:Reconstruction_different_prior}). The results reveal that GOOD provides even stronger protection of client data privacy against weaker adversaries, with the median SSIM consistently remaining \textbf{below 5\%}. 
Overall SSIM performance is very consistent in the CNN architecture, with the most variation coming from the Euclidean distance based attack when run with the FMNIST dataset. Interestingly, the SSIM score tends to lower as the priors become stronger when using the CNN architecture. The performance is also very consistent in the ResNet18 architecture, with variation being highest when run with the UTKface dataset. The MSE scores are also very consistent within both architectures, with the MSE scores being higher and having more variation within the CNN, indicating the reconstruction was less effective there. %
GOOD is effective in privacy protection against diverse adversarial scenarios with various levels of prior knowledge. It maintains consistently strong performance against reconstruction in terms of SSIM across all datasets and architectures. 
 
{\color{black} DP-SGD has strong performance in terms of MSE and SSIM, effectively protecting client privacy. For all attacks, the SSIM score is also under 5\%, with the medians being even lower than those found in the GOOD results. The only exception is the GradInversion ResNet18 results with the FMNIST dataset. The MSE results are generally higher than those seen with GOOD, with a significantly strong performance against Euclid. The other MSE results have higher medians but much more variation than those seen in GOOD. %
Overall, while stronger in some cases, DP-SGD's performance is comparable to our GOOD approach. These results show that GOOD can provide protection nearly on par with DP-SGD without significantly impacting a model's utility.

Note that while we compare GOOD's performance in preventing privacy violations via reconstruction attacks, it does not provide worst-case guarantees like the DP-SGD approach. Although early experiments are encouraging, future work will include a more comprehensive theoretical and empirical assessment and comparison.}

\subsection{System Scalability Analysis}
\label{sec:eval-scalability}
{\bf Deployment Setup:} To evaluate the scalability of our approach, we conducted experiments in two settings: {\it (i)} a one-to-one setup, where a single client connects to a single aggregator, and {\it (ii)} a many-to-one setup, where multiple clients simultaneously connect to a single aggregator. In the one-to-one setup, we tested the client application on multiple devices with varying specifications: a server-grade desktop, a consumer-grade desktop, and a Raspberry Pi 3. The server-grade desktop featured an Intel(R) Xeon(R) W-2245 CPU clocked at 3.90 GHz with 128 GiB of RAM, the consumer-grade desktop was powered by an AMD Ryzen 9 7900X 12-Core Processor running at 4.70 GHz with 64 GiB of RAM, and the Pi 3 runs an ARM Cortex-A53. The aggregator was deployed on a standard DC16s v3 Microsoft Azure Virtual Machine (VM), equipped with 16 virtual CPUs and 128 GiB of memory.
In the second setup, the client and server applications were deployed on Microsoft Azure standard D16s v5 VMs, each equipped with 16 vCPUs and 64 GiB of memory. Using these machines, we experimented with incrementally increasing the number of simultaneous connections to the aggregator and measuring the average latency. The aggregator application was hosted on a VM of similar capability, for consistency.

\begin{figure*}[ht]
\centering
    \begin{minipage}[t]{0.3\textwidth}
        \centering
        \includegraphics[height=4cm,width=\linewidth]{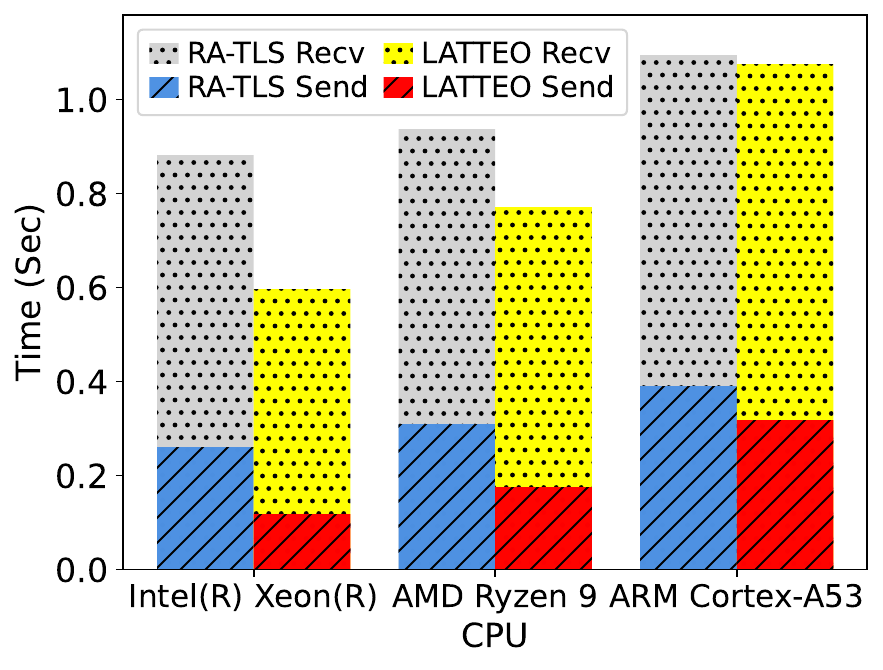}
        \vspace{-0.3in}
        \caption{Comparison of the communication between the client and server in RA-TLS vs \sysname.} 
        \label{fig:LatOnboarding}
    \end{minipage}%
    \hspace{0.02\textwidth}
    \begin{minipage}[t]{0.3\textwidth}
        \centering
        \includegraphics[height=4cm,width=\linewidth]{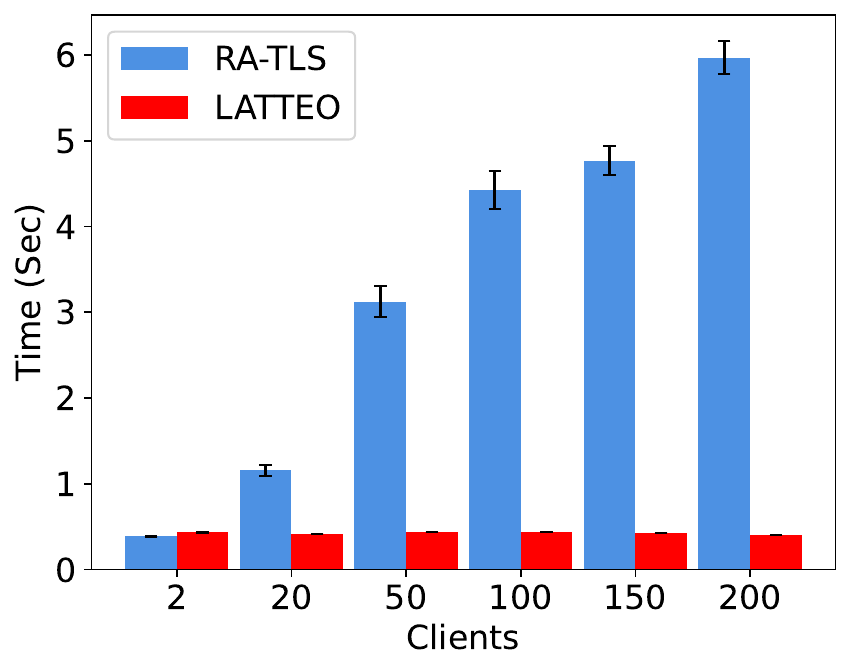}  
        \vspace{-0.3in}
        \caption{Average time for the initial connection across different number of simultaneous clients.}
        \label{fig:ScaleTest}
    \end{minipage}%
    \hspace{0.02\textwidth}
    \begin{minipage}[t]{0.3\textwidth}
        \centering
        \includegraphics[height=4cm,width=\linewidth]{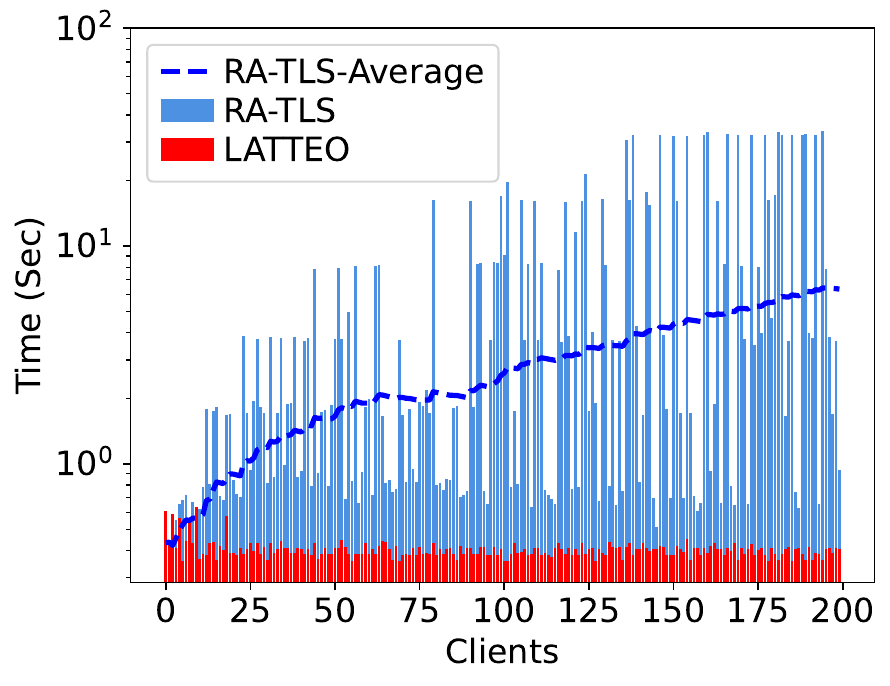}  
        \vspace{-0.3in}
         \caption{Communication latency as the number of simultaneous connections increases (log scale).}  
        \label{fig:First1000}
    \end{minipage}
    \vspace{-0.1in}
\end{figure*}

\noindent
{\bf Results Analysis:} For system scalability, we began by evaluating the end-to-end latency of our system. Figure~\ref{fig:LatOnboarding} provides a breakdown of the time between the client and server in a one-to-one setup, detailing the communication latency which encompasses the elapsed time for cryptographic operations. 
In all cases, except for the Pi 3 (ARM Cortex-A53), there is a noticeable reduction in overall latency despite the additional cryptographic operations introduced by \sysname. Notably, the time spent on these operations has minimal impact on the overall communication latency. Table~\ref{table:times} lists the latency breakdown of \sysname operations. {\color{black} MABE's encryption operation totals around 28.01~ms in the worst case, accounting for at most 4.7\% of the total communication time. The symmetric key encryption operation takes less than 3~ms except when on the ARM processor where it takes 14~ms accounting for 1.3\% of the total communication time.} These times are negligible when compared to the total time taken by the communication process, and as seen in Figure~\ref{fig:LatOnboarding} do not affect the overall performance of \sysname.

An interesting trend is that the latency gap between \sysname and RA-TLS decreases as the available RAM of the device decreases. This can be attributed to the use of MsQuic, which spawns background threads that may strain memory-limited devices. These results indicate that on devices with sufficient resources, \sysname significantly reduces user-perceived latency compared to TLS-TCP and, at worst, introduces no additional latency. 

\begin{table}[t]
\small
\centering
\caption{\label{table:times} The breakdown of the client’s perceived latency, averaged across multiple runs.}
\label{results} 
\vspace{-0.1in}
\begin{tabular}{ccccc}
\toprule
\hline
    {\bf Processor}     & {\bf Total} & {\bf Comm}   & {\bf Sym. Key}  & {\bf MABE}     \\ \midrule
    Intel(R) Xeon(R)    & 592.87 ms     & 565.37 ms     & 2.75 ms       & 28.01 ms \\ \hline
    AMD Ryzen 9         & 784.4 ms     & 758.8 ms      & 0.85 ms       & 10.85 ms \\ \hline
    ARM Cortex-A53      & 1089.06 ms     & 1048.48 ms    & 14 ms         & 12.7 ms  \\ 
\hline
\bottomrule
\end{tabular}
\vspace{-0.2in}
\end{table}

\begin{figure*}[h]
\centering
    \begin{minipage}[t]{0.31\textwidth}
        \centering
        \includegraphics[height=4cm, width=\linewidth]{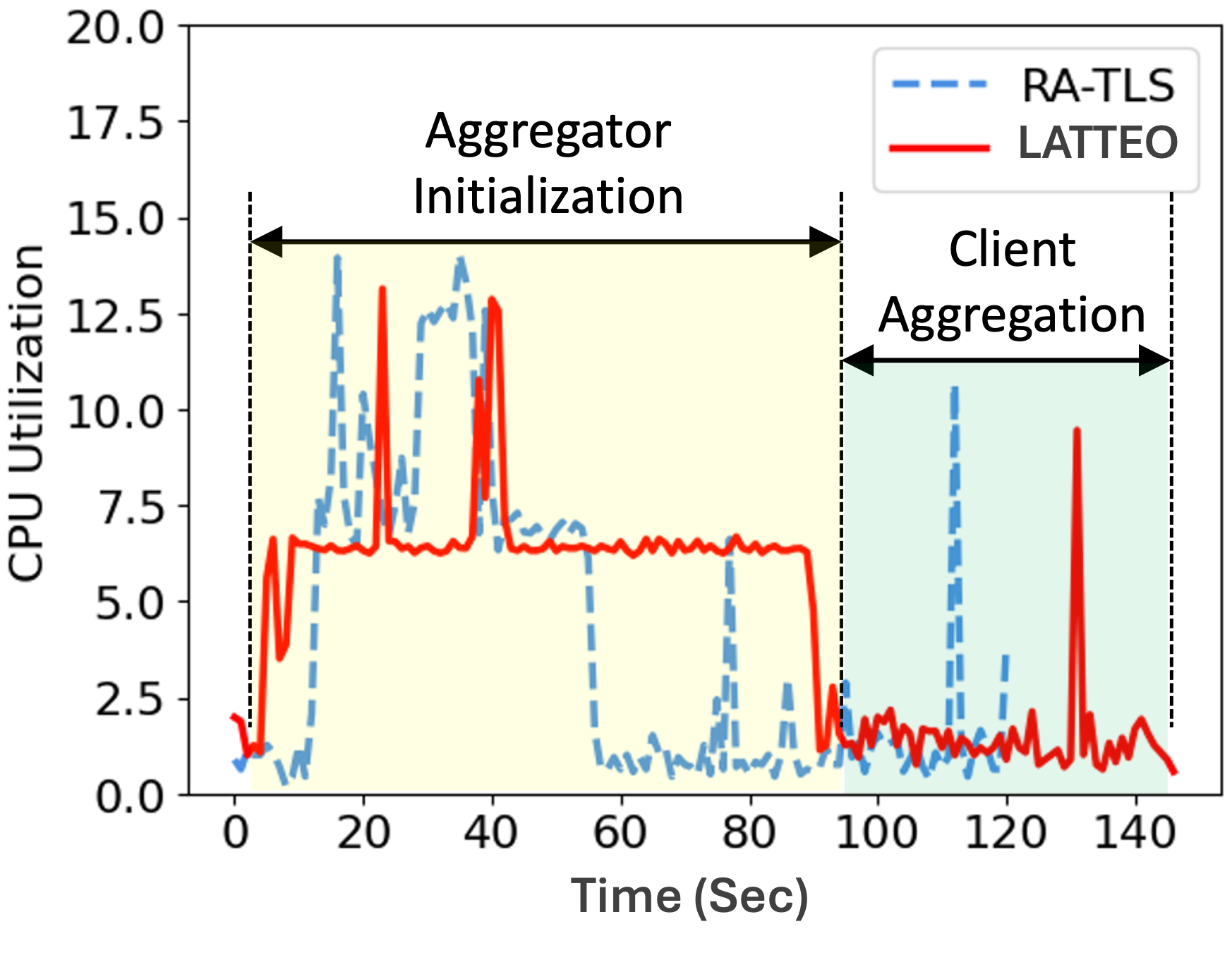}   
        \vspace{-0.25in}
        \caption{Server CPU Utilization}   
        \label{fig:CPUServer}
    \end{minipage}%
    \hspace{0.01\textwidth}
    \begin{minipage}[t]{0.31\textwidth}
        \centering
        \includegraphics[height=4cm,width=\linewidth]{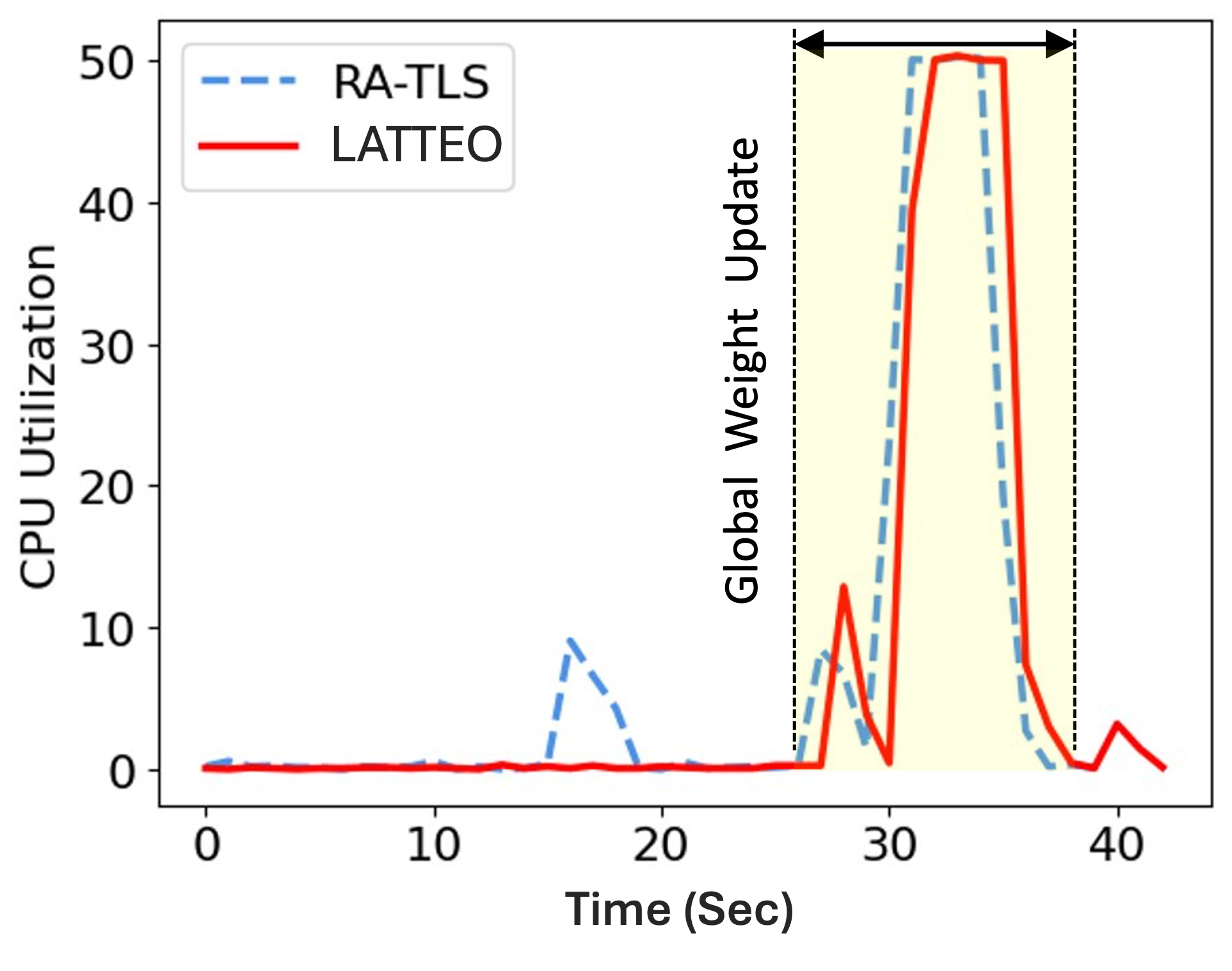}  
        \vspace{-0.25in}
        \caption{Client CPU Utilization}   
        \label{fig:CPUClient}
    \end{minipage}%
    \hspace{0.01\textwidth}
    \begin{minipage}[t]{0.31\textwidth}
        \centering
        \includegraphics[height=4cm,width=\linewidth]{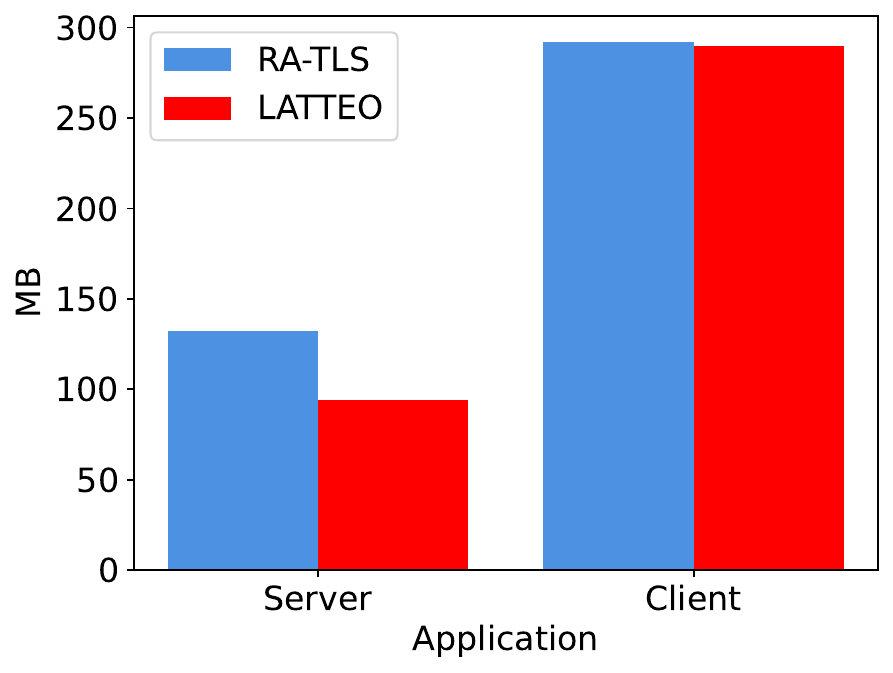} 
        \vspace{-0.25in}
        \caption{Memory Utilization}   
        \label{fig:Memory}
    \end{minipage}
    \vspace{-0.1in}
\end{figure*}

We employed the many-to-one setup to evaluate how increasing the number of active connections to the server affects communication for clients. In this series of experiments, we progressively increased the number of active clients attempting to register and retrieve the global weights from the aggregation server. For instance, in our experiments with 200 simultaneous clients, there were always 200 active clients competing to register with the aggregator at any given time. This was achieved by launching a new client instance as soon as another completed its registration. For each experiment, we measured the average time required for a client to complete the registration process.
Figure~\ref{fig:ScaleTest} illustrates the average latency as the number of clients increases from 2 to 200. Initially, with just two simultaneous connections, RA-TLS and \sysname exhibit comparable performance. However, as the number of simultaneous connections grows, RA-TLS shows a significant increase in average latency, whereas \sysname consistently maintains low latency. 
Specifically, with 200 concurrent clients' connections, \textbf{RA-TLS demonstrates an average latency nearly 1500\% higher than that of \sysname}.

To further investigate the root cause of RA-TLS’s behavior, we analyzed per-connection latency as the number of clients increased (Figure~\ref{fig:First1000}). This latency was calculated as the total time a client takes to connect to the server, including the re-establishment of timed-out connections. It is evident that the increased latency in RA-TLS is directly correlated with the growing number of connection timeouts, particularly when the increasing number of concurrent connections, between 100 to 200 connections, overwhelms the RA-TLS server. Note that we did not consider any extra load balancing service for either of the approaches in an attempt to replicate the edge environment.
These results demonstrate that the \sysname framework scales effectively and is better suited to handle many simultaneous users than the RA-TLS framework.

\subsection{System Cost Analysis}

To evaluate the cost incurred by our system, we analyzed the CPU utilization and memory footprint of both \sysname and RA-TLS.
Figure~\ref{fig:CPUServer} shows the CPU utilization of the aggregator application over time for both \sysname and RA-TLS. During the initialization phase, both frameworks exhibit a CPU utilization of approximately 6\%. While their performance during initialization is relatively similar, RA-TLS shows more frequent spikes, whereas \sysname has a startup time about 30 seconds longer. This discrepancy can be attributed to \sysname's use of MsQuic, which initializes a series of threads for communication, all of which must be provisioned within the SGX enclave. During the client aggregation phase, both frameworks exhibit similar CPU utilization.
A comparison of client-side CPU utilization on the server-grade desktop, shown in Figure~\ref{fig:CPUClient}, indicates that both frameworks have nearly identical utilization profiles, although, the CPU usage increases to approximately 50\% for a few seconds while processing the global model and cryptographic operations have minimal impact on CPU usage.

The maximum memory consumption of the aggregator and client applications, as shown in Figure~\ref{fig:Memory}, indicates that both frameworks demonstrate similar performance. However, the RA-TLS implementation shows slightly higher memory usage for the server application. This difference can be attributed to the additional state management required by TLS connections use in the RA-TLS.

%% file: sec02.tex
\section{Related Work}
\label{sec02}
Despite the privacy advantages of the FL paradigm, it remains vulnerable to various privacy and security attacks\cite{nasshohou2019comprehensive,zhuLiGu2024evaluating,shostrmarc2017membership,melsondec2019exploiting,picromveg2023Perfectly,pasfraate2022Eluding}. A significant attack vector in distributed learning is data reconstruction~\cite{yanGeXia2023using, BoeDziSchu2023when,ZhaShaElk2024Large-scale}

These attacks undermine one of the primary motivations for adopting FL: client privacy. To address these vulnerabilities, various security enhancements have been proposed, including sharing only portions of the gradients during updates and employing participant-level differential privacy~\cite{ShoRezShm2015Privacy,weilima2023personalized,Hewancai2024clustered,HuGuoGon2023Federated,liuligao2023privacy-encoded}, albeit often at the cost of model accuracy. Another defense strategy enhances client confidentiality through a Double-Masking protocol for gradients and counters adversarial aggregators by enforcing proof of correctness during aggregation~\cite{xuliliu2020verifynet}.
Numerous novel defenses against reconstruction attacks have also been proposed. For instance, the approach in~\cite{NaHyeJun2022Closing} introduces a low-cost method for obscuring gradients, while another approach incorporates Trusted Execution Enclaves (TEEs) into a synchronous FL (SyncFL) environment, isolating sensitive components of the aggregation process from the rest of the operating system~\cite{Caozhazha2024SRFL}.

These solutions were designed for synchronous FL (SyncFL) systems, with limited consideration for asynchronous FL (AsyncFL) systems. Due to the intrinsic assumption that clients and aggregators operate in a synchronized manner, many of these solutions are not directly applicable to AsyncFL. As a result, they often fail to provide sufficient security or achieve convergence in asynchronous environments.
Nonetheless, some defenses have been developed specifically for AsyncFL. For instance, certain approaches focus on detecting and preventing malicious clients~\cite{FanLiuGon2022AFLGuard,tiacheyu2021towards,liuyuzon2024Delay}. Others propose modifications to the AsyncFL system to enhance speed while maintaining some of the privacy and security levels of SyncFL~\cite{ngumalzha2022federated}. Nonetheless, these solutions still have gaps, particularly in defending against malicious clients and dishonest aggregators while preserving system efficiency--gaps that we addressed.

%% file: full-formalsecurity.tex
\section{\sysname Security Analysis}
\label{sec:full-sec-analysis} 
\subsection{Formal Security Analysis}
We now provide a formal analysis of \sysname in the Universal Composability (UC) Framework~\cite{Canetti01}. The notion of UC security is captured by the pair of definitions below:

\begin{definition}{(UC-emulation~\cite{Canetti01})}
\label{def:ucdef1}
Let $\pi$ and $\phi$ be probabilistic polynomial-time (PPT) protocols. We say that $\pi$ UC-emulates $\phi$ if for any PPT adversary $\adv$ there exists a PPT adversary $\simu$ such that for any balanced PPT environment $\env$ we have
$$ \mathrm{EXEC}_{\phi, \simu, \env}  \approx \mathrm{EXEC}_{\pi, \adv, \env} $$
\end{definition}
 
\begin{definition}{(UC-realization~\cite{Canetti01})}
\label{def:ucdef2}
Let $\mathcal{F}$ be an ideal functionality and let $\pi$ be a protocol. We say that $\pi$ UC-realizes $\mathcal{F}$ if $\pi$ UC-emulates the ideal protocol for $\mathcal{F}$.
\end{definition}

We define an ideal functionality, $\fdctc$, consisting of five independent ideal functionalities, $\fregister, \fresp, \ftee%
, \fsmt, \fsig$%
.
The parties that interact with the ideal functionalities are the members of sets of service providers, $\mathbb{SP}$, edge servers with enclave execution capability, $\mathbb{E}$, %
and users $\mathbb{U}$. We assume that each member of the sets has a unique identifier.
We assume that $\fdctc$ maintains an internal state that is accessible at any time to all the ideal functionalities, specifically  %
tables,  $\stable$, $\spidapptable$, $\ABEetable$, $\abesktable$, $\symmktable$, and $\symmetable$. %
$\stable$ contains service providers information that an edge server provides services on behalf of, such as the set of attributes used by the enclaves and the service provider's ABE public key. $\spidapptable$ contains encrypted apps and information stored by each service provider that are

\begin{figure}[H]
\begin{mdframed}
\begin{center}
\textbf{Functionality} $\fregister$
\end{center}

\begin{enumerate}%

\item \textbf{system setup:} on receiving request $(systemsetup,\mathbb{SP},\lambda,sid)$ from $\env$, $\fregister$ sets the system parameter $\lambda$, the set of service providers in the system  as $\mathbb{SP}$, set of edge servers $\mathbb{E}$, and set of users as $\mathbb{U}$. %

\item \textbf{ABE setup:} on receiving request $(\abesetup,$ $\spid, [a1,\dots,$ $an],$ $sid)$ from a $SP \in \mathbb{SP}$ identified by $\spid$, $\fregister$  %
picks $mpk_{\spid} \sample \{0,1\}^{\lambda}$ and adds 
$(\spid,[a1,\dots, an],$ $mpk_{\spid})$ 
to $\stable$. It returns $(\spid,[a1,\dots, an],mpk_{\spid})$ to $\simu$ and $\adv$. %

\item \textbf{ABE keygen:} On receiving request $(\abekeygen,[ai],sid)$ from a $SP \in \mathbb{SP}$ identified by $\spid$, $\F{register}$ checks if $(\spid,[\dots ai,\dots],mpk_{\spid})$ exists in $\stable$, if so, it picks $sk_{a1} \sample \{0,1\}^{\lambda}$, stores $(\spid, sk_{a1},a1)$ in $\abesktable$, and returns $sk_{a1}$ to $\spid$.

\item \textbf{ABE encryption:} On receiving request $(\abeenc,policy,mpk_{\spid},m,sid)$ from user $u$, $\F{register}$ checks if a tuple $(\cdot,[a1,\dots, an],mpk_{\spid})$ exists in $\stable$ and that attributes in $policy$ also exist in $[a1,\dots,an]$. If not, return $\bot$, else it picks $c \sample \{0,1\}^{\lambda}$, and adds  $(mpk_{\spid},policy,c,m)$ to $\ABEetable$ and returns $c$ to $u$, $\adv$, and $\simu$. 

\item \textbf{ABE decryption:} On receiving request $(\abedec,[sk_{aj},\dots,sk_{ak}],policy,c,sid)$ from user $u$, $\F{register}$ checks if a tuple $(mpk_{\spid},policy,c,\cdot)$ exists in $\ABEetable$, that attributes $[sk_{aj},\dots,sk_{ak}]$ satisfy $policy$, and  there exist tuples $(\cdot,sk_{ai},ai)$ in $\abesktable$ for each $sk_{ai} \in [sk_{aj},\dots,sk_{ak}]$. %
If previous checks pass then it retrieves the tuple $(mpk_{\spid},policy,c,m)$ from $\ABEetable$ and returns $m$ to $u$, else return $\bot$.

\item \textbf{Symmetric key generation:} On receiving request $(\symmkeygen,sid)$ from user $u$, $\F{register}$ picks $k \sample \{0,1\}^{\lambda}$, and adds  $(k)$ to $\symmktable$ and returns $k$ to user and $\simu$. %

\item \textbf{Symmetric key encryption:} On receiving request $(\symmenc,k,m,sid)$ from user $u$, if $k$  exists in $\symmktable$, $\F{register}$ picks $c \sample \{0,1\}^{\lambda}$, and adds  $(k,c,m)$ to $\symmetable$ and returns $c$ to $u$, $\adv$, and $\simu$. Else return $\bot$. %

\item \textbf{Symmetric key decryption:} On receiving request $(\symmdec,k,c,sid)$ from user $u$,  $\F{register}$ checks if a tuple $(k,c,\cdot)$ exists in $\symmetable$, if so it retrieves the tuple $(k,c,m)$ and returns $m$ to $u$, else return $\bot$.%

\item \textbf{Service provider setup:} on receiving request $(\mathrm{SP\_setup},$ $\spid,$ $encCoorApp,K_C,$  $encAggApp,K_A,$$bA, attestPrim,sid)$ from a $SP \in \mathbb{SP}$ identified by $\spid$, $\fregister$ adds $(\spid, encCoorApp, K_C$ $encAggApp, K_A,bA, spec)$ to $\spidapptable$.

\end{enumerate}

\end{mdframed}
\vspace{-0.15in}
\caption{Ideal functionality for Service Registration}
\label{fig:ucregister1}
\end{figure}

\noindent sent to edge servers when they register for a service. $\abesktable$ stores the secret keys for the ABE attributes generated by the different service providers and $\ABEetable$ stores the ABE encrypted messages, thus helping us simulate ABE functionality in the ideal world. $\symmetable$ and $\symmktable$ help realize symmetric key encryption operations in the ideal world by storing ciphertexts and symmetric keys respectively.  %
Lastly, $\etable$ helps keep track of edge servers registered with service providers to provide their services.

We note that the service providers in \sysname are trusted entities and do not act maliciously and that the users in the system are not authenticated. Any user in the system has access to all service providers' public keys and can encrypt an aggregation request for an aggregation enclave. The edge servers running the coordinator enclave and the aggregation enclave are honest but curious entities and will try to undermine the system security by accessing the user data that the enclaves use. We note that user authentication can be easily achieved in the application of \sysname by plugging in off-the-shelf protocols such as token-based (Java Web Token) and certificate-based authentication schemes.
We now briefly describe the design of our ideal functionalities.

\addtocounter{figure}{-1}

\begin{figure}[h!]
\begin{mdframed}
\begin{center}
\textbf{Functionality} $\fregister$
\end{center}

\begin{enumerate}\addtocounter{enumi}{9}%

\item \textbf{Get all public keys:} on receiving request $(retrieve\_mpks, sid)$ from user $u$,  $\fregister$ retrieves all tuples $(\cdot,[a1 \dots an],mpk_{\spid})$ from $\stable$, generates a list $[(mpk_{\spid},[a1,\dots]),\dots]$ %
with each $mpk_{\spid}$ in the list corresponding to an $SP \in \mathbb{SP}$ and the corresponding attribute list for the $SP$. $\fregister$ sends list to $u$, $\adv$, and $\simu$.

\item \textbf{Coordinator edge server registration:} On receiving request $(registerCoorEnc$$,$$\spid,$$e,$$sid)$ from $e$, $\fregister$ checks if a tuple $(\spid,$ $encCoorApp,$ $\cdot,$$encAggApp,$$\cdot,$$bA,$$attestPrim)$ exists in $\spidapptable$, if true, sends $(encCoorApp,$ $bA)$ to $e$, $\simu$, and $\adv$, else return $\bot$. 

\item \textbf{Aggregation edge server registration:} %
On receiving request $(registerAggEnc,$ $\spid,$ $pid,$ $sid)$ from $e$, $\fregister$ retrieves $(\spid, encCoorApp,\cdot,encAggApp,\cdot,bA,attestPrim)$ from $\spidapptable$,  %
and sends $(encAggApp,bA)$  to $e$, $\simu$, and $\adv$, else return $\bot$.

\end{enumerate}
\end{mdframed}
\vspace{-0.15in}
\caption{Ideal functionality for Service Registration}
\label{fig:ucregister2}
\end{figure}

$\underline{\fregister}$: The $\fregister$ functionality shown in Figure~\ref{fig:ucregister1} handles the system setup, setup of service providers, coordinator and aggregation enclave server registrations, and functions to support attribute-based encryption functionality and symmetric key encryption functionality. %
$\env$ first initializes the system by sending the tuple $(systemsetup,\mathbb{SP},\lambda,sid)$ to $\fregister$. This initializes the set of entities ($\mathbb{SP}$) in the system that act as service providers, $\mathbb{E}$ as the set of edge servers, and the set of users in the system $\mathbb{U}$. 
Each service provider in the system, $SP \in \mathbb{SP}$,  contacts $\fregister$ by sending a tuple  $(\abesetup,$ $\spid, [a1,\dots,$ $an],$$sid)$  where $\spid$ denotes the unique identifier of $SP$ and $[a1,\dots,$ $an]$ denotes the list of attributes for which the $SP$ wants to generate ABE keys corresponding to the services the $SP$ is offering. %
$\fregister$ stores the information in $\stable$ and returns a unique $mpk_{\spid}$ to $\spid$. This function allows for overwriting a previous tuple in $\stable$ to account for $SP$s changing the attribute set or services they offer. %
$SP$s send $(\abekeygen,[ai],sid)$ tuple to $\fregister$ to generate secret keys associated with the attribute $ai$. $\fregister$ checks $\stable$ for the existence of the attribute $ai$ before generating the key and returning it to $SP$ identified by $\spid$. Any user in the system can call $\fregister$ with tuple $(\abeenc,policy,$ $mpk_{\spid},$ $m,sid)$ to get a message $m$ encrypted using public key $mpk_{\spid}$ and an ABE policy ($policy$) containing valid attributes of $SP$ identified by $\spid$. Any user in the system with valid secret keys for attributes in a give encryption policy can call $\fregister$ with tuple $(\abedec,[sk_{aj},\dots,sk_{ak}],policy,c,sid)$. $\fregister$ checks that the supplied secret keys are valid for the given policy and have been generated previously. If all the checks pass then $\fregister$ retrieves the decrypted message from $\ABEetable$ and returns it to the calling entity. $\fregister$ also provides symmetric key generation, encryption, and decryption functions. $\fregister$ creates symmetric keys and returns them to the calling user after storing them in the $\symmktable$. For encryption, a user calls $\fregister$ with $(\symmenc,k,m,sid)$, if the key $k$ exists in the $\symmktable$, then $\fregister$ stores a ciphertext $c$ corresponding to $m$ in $\symmetable$ and returns $c$ to the calling user. Any user in the system with cipher text $c$ and a valid key $k$ corresponding to $c$ can call the decryption function of $\fregister$ by sending the tuple $(symmdec,k,c,sid)$. $\fregister$ checks for the existence of a tuple containing $c$ and $k$ in $\symmetable$, and if found, returns the corresponding plaintext message $m$, thus providing the symmetric key decryption functionality.
\par 

After creating their corresponding ABE and symmetric keys, service providers in the system store the encrypted coordinator and aggregator applications with $\fregister$ by calling $(\mathrm{SP\_setup},$ $\spid,$ $encCoorApp,K_C,$  $encAggApp,K_A,$ $bA,$ $attestPrim,$ $sid)$. The encrypted apps and $bA$ will be sent to edge servers upon their registration as coordinator or aggregation enclaves. The symmetric encryption keys $K_C$ and $K_A$ are accessed by $\ftee$ when it decrypts the $encCoorApp$ and $encAggApp$ inside the enclave, respectively. \\$attestPrim$ is information that is used by the service providers and coordinator enclaves in the real world for remote attestation of provisioned enclaves and verifying that they were set up correctly before sending decryption keys for the applications and user data.  
Any user in the system can call $\fregister$ by sending tuple $(retrieve\_mpks, sid)$ to retrieve the set of ABE public keys for all $SP$s in the system as well as the attributes associated with the public keys. 
This models the fact that all service providers' public keys are available to any user in the system in the real world. 
An edge server $e$ in the system can register for providing coordinator enclave service by sending the tuple $(registerCoorEnc,\spid,e,sid)$ to $\fregister$. $\fregister$ retrieves the encrypted coordinator app $encCoorApp$ and the boot app $bA$ and sends it to $e$. %
An edge server $e$ in the system can register for providing aggregation enclave service by sending the tuple $(registerAggEnc,\spid,e,sid)$ to $\fregister$. $\fregister$ retrieves the encrypted aggregation app, $encAggApp$ and the boot app $bA$ and sends it to $e$. %

\begin{figure}[h]
\vspace{-0.0in}
\begin{mdframed}
\begin{center}
\textbf{Functionality} $\fresp$
\end{center}
\begin{enumerate}%

\item \textbf{User request:} On receiving request $(userRequest,C1,C2,Cert_{u},e,u,sid)$ from user $u \in \mathbb{U}$, $\fresp$ returns $\bot$ to $u$ if no such $e$ exists, else forwards request to edge server $e$, $\simu$, and $\adv$. %

\item \textbf{Edge server response:} On receiving message $(serverResponse,C3,u,sid)$ from edge server $e$, $\fresp$ forwards request to user $u$, $\simu$, and $\adv$.

\end{enumerate}
\end{mdframed}
\vspace{-0.15in}
\caption{Ideal functionality for Responding to Requests}
\label{fig:ucresponse}
\end{figure}

$\underline{\fresp}$: %
The $\fresp$ functionality shown in Figure~\ref{fig:ucresponse} handles a service request from a user identified by $u$. When $u$ submits a request to $\fresp$ for aggregation service provided by an edge server $e$, it sends a request containing $(userRequest,C1,C2,Cert_{u},e,u,sid)$. %
$C1$ and $C2$ in the request are encrypted by the user using symmetric key encryption and attribute based encryption, respectively, as in the real world. $e$ helps the UC functionality forward the request to the appropriate edge server running the aggregation enclave.
Once $\fresp$ receives the request from the user, it forwards the tuple to $e$, $\adv$, and $\simu$. If no such server $e$ exists then $\bot$ is returned to $u$. 
After processing a user $u$'s request, an edge server $e$ can call $\fresp$ by sending the tuple $(serverResponse,C3,u,sid)$, where $C3$ is a symmetric key encrypted data meant for user $u$. $\fresp$ forwards the request to user $u$, $\simu$, and $\adv$.

$\underline{\ftee}$: Figure~\ref{fig:ftee} represents the ideal functionality for a secure enclave as described in Pass \kETAL~\cite{pass2017formal}. $\ftee$ is used by edge servers to launch the $BootApp$, represented by $bA$ in the ideal world by calling $(install,idx,\prog)$, where $\prog$ is $bA$. The $BootApp$ executes the coordinator app as well as the aggregation app within the enclave. The steps executed within secure enclaves described in the protocols in Section~\ref{sec05} are executed inside the $\ftee$ functionality in the ideal world. %
$\ftee$ also encrypts the user data $C3$ before returning it in the form of $\outp$ to the calling server $e$ which would have executed the aggregation enclave application when calling $\ftee$ using $install$ function. $\ftee$ can internally access all the tables maintained by $\fregister$ which will allow $bA$ running inside $\ftee$ to retrieve decryption keys from $\symmetable$ and decrypt $encCoorApp$ and $encAggApp$ as needed.

\begin{figure}[h]
\begin{mdframed}
\begin{center}
\textbf{Functionality} $\ftee$
\end{center}

\textbf{initialize:}
On receiving ($ini$,$\sid$), set $mpk,msk$ = $\sum.\KeyGen(1^{\lambda})$, $T = \emptyset$. %

\textbf{public query:}
on receiving $\getpk()$ from some $\party$, send $mpk$ to $\party$.
\vspace{0.1in}
\begin{mdframed}
\begin{center}
Enclave operations
\end{center}

\textbf{public query:} on receiving $(install,idx,\prog)$ from some $\party \in \reg$, if $\party$ is honest, assert $idx == sid$, generate nonce $\eid \in \{ 0,1 \}^{\lambda}$, store $T[ \eid,\party] = (idx, \prog,\overrightarrow{0})$, send $\eid$ to $\party$.

\textbf{public query:} on receiving $(resume, \eid, \inp)$ from some $\party \in \reg$,\\
let $(idx, \prog, \mem) = T[\eid,\party]$, abort if not found.\\
let $(\outp,\mem) = \prog(\inp,\mem)$, update $T[\eid,\party] = (idx,\prog,\mem)$\\
let $\sigma = \sum.Sig_{msk}(idx, \eid, \prog, \outp)$, and send $(\outp, \sigma)$ to $\party$.
\end{mdframed}

\end{mdframed}
\caption{Ideal functionality for the Secure Enclave~\cite{pass2017formal}}
\label{fig:ftee}
\end{figure}

%% file: appendix.tex
\subsection{Proof}

We now prove Theorem~\ref{thm:uc}.

\begin{proof}
We give a series of games, each of which is indistinguishable from its predecessor by a PPT $\env$.

$\gzero$: This is the same as the real-world \sysname. $\env$ interacts directly with \sysname and $\adv$.

$\gone$: $\simu$ internally runs $\adv$ and simulates the secure and authenticated channels functionality $\fsmt$. 
    \begin{lemma}
    \label{lem:lem1}
    For all PPT adversaries $\adv$ and PPT environments $\env$, there exists a simulator $\simu$ such that 
    $$ \execgzero \approx \execgone
    $$ \end{lemma}
    The two games are trivially indistinguishable since $\simu$ just executes the simulator for $\fsmt$.

$\gtwo$: $\simu$ communicates with the honest parties and $\adv$, and simulates the protocols of \sysname with the help of $\fdctc$. $\adv$ can corrupt any user or $EC$ at any point in time by sending a message ``corrupt'' to them. Once an entity is corrupted, all their information is sent to $\adv$ and all further communication to and from the corrupted party is routed through $\adv$. We now state and prove the following lemma:

\begin{lemma}
\label{lem:lem2}
For all PPT adversaries $\adv$ and PPT environments $\env$, there exists a simulator $\simu$ such that  
$$\execgone \approx \execgtwo$$
\end{lemma}

 After system setup, $SP$s can independently call the ABE setup function in $\fregister$ to set up their corresponding public keys and attributes. The attributes act as descriptors for specific services aggregation enclaves and coordinator enclaves will provide to users and these attribute-specific keys will be accessible to the enclave applications through RA-TLS connection in the real world, which is modeled in the ideal world by $\ftee$ accessing the $\abesktable$. $SP$s would then create their individual symmetric keys $K_A$ and $K_C$ by calling the $\symmenc$ function in $\fregister$ for encrypting their aggregator and coordinator apps, respectively. $SP$s will call the $\symmenc$ function and pass their coordinator and aggregator apps to generate $encCoorApp$ and $encAggApp$. Service providers call the $\mathrm{SP\_setup}$ function of $\fregister$ to store their encrypted applications and $BootApp$ in $\spidapptable$. 

Edge server entities in $\mathbb{E}$ which will provide coordinator services can call the $registerCoorEnc$ function of $\fregister$ to receive the $BootApp$ ($bA$) and the $encCoorApp$ for a specific service provider $SP \in \mathbb{SP}$ identified by $\spid$. 
Edge servers that will provide aggregation services can call the $registerAggEnc$ function of $\fregister$ to receive the $BootApp$ ($bA$) and the $encAggApp$ for a specific service provider $SP \in \mathbb{SP}$ identified by $\spid$. 
After receiving the $bA$ and the encrypted enclave app, the edge servers will initialize the bootapp using $\ftee$. In the real world, the enclave connects to the service provider to participate in remote attestation to verify that it has been initialized correctly before receiving the symmetric key $K_C$ or $K_A$ to decrypt the encrypted coordinator app or aggregator app, respectively. In the ideal world, $\ftee$ has access to the internal state maintained by all other functionalities, and hence, $\ftee$ can retrieve the $K_C$ and $K_A$ stored by the service provider in the $\spidapptable$. All the steps inside the aggregation and coordinator enclave in the ideal world as executed by $\ftee$ and the edge server has no access to the internal state of the $\ftee$. When an aggregation enclave is initialized, the verification in the real world that occurs between the aggregation enclave and coordinator enclave is handled within $\ftee$ as well, which models the secure communication between secure enclaves in the real world since no user interaction is involved during this stage.

When a user $u$ in the system needs to create a request for the aggregation service, they will first access the ABE public keys for the specific service provider. In the real world, this data is publicly accessible and this is modeled in the ideal world by calling the $retrieve\_mpks$ function which returns the public keys as well as the attributes for all the $SP$s in the system. The user $u$ generates a symmetric key $K_U$ using $\fregister$ and calls $\symmenc$ to encrypt their data using $K_U$, which results in $C_1$. The user further encrypts $K_U$ using the $\abeenc$ function of $\fregister$ using the service provider public key and the attribute associated with the aggregation service, resulting in $C_2$. Both the encrypted data are then sent to an aggregation server $e$ by calling the $userRequest$ function provided by $\fresp$. The edge server $e$ has to load the encrypted data into the enclave, which in the ideal world requires sending the data to $\ftee$ which is currently executing the $AggApp$ for $e$. 
$\ftee$ internally decrypts $C_2$ using the $\abedec$ function and then uses the decrypted key $K_U$ to access user data in $C_1$. The result of the computation is encrypted by $\ftee$ using $K_U$  and sent to the edge server $e$. 
At this point, $e$ calls the $serverResponse$ function of $\fresp$ and forwards the encrypted response from the enclave to $\fresp$ who forwards the response to $u$. 
Since $u$ has access to $K_U$, it calls the $\symmdec$ function in $\fregister$ to decrypt the response and concludes the protocol.

We note that this proof only accounts for the communication security aspect of \sysname and does not handle reconstruction attacks the user can mount given its access to the response from the aggregation service. The ideal world adversary can only try to access the user data before it is passed to the enclave and the responses from the aggregation enclave. Since this information is protected with symmetric key encryption and attribute-based encryption, as long as the security guarantees of the encryption schemes hold and the trusted enclave is secure, the adversary cannot access the user data which is only available to the enclaves and the individual users in the system.

    \qed

\end{proof}

%% file: paper.bbl

\begin{thebibliography}{55}


\ifx \showCODEN    \undefined \def \showCODEN     #1{\unskip}     \fi
\ifx \showDOI      \undefined \def \showDOI       #1{#1}\fi
\ifx \showISBNx    \undefined \def \showISBNx     #1{\unskip}     \fi
\ifx \showISBNxiii \undefined \def \showISBNxiii  #1{\unskip}     \fi
\ifx \showISSN     \undefined \def \showISSN      #1{\unskip}     \fi
\ifx \showLCCN     \undefined \def \showLCCN      #1{\unskip}     \fi
\ifx \shownote     \undefined \def \shownote      #1{#1}          \fi
\ifx \showarticletitle \undefined \def \showarticletitle #1{#1}   \fi
\ifx \showURL      \undefined \def \showURL       {\relax}        \fi
\providecommand\bibfield[2]{#2}
\providecommand\bibinfo[2]{#2}
\providecommand\natexlab[1]{#1}
\providecommand\showeprint[2][]{arXiv:#2}

\bibitem[Anati et~al\mbox{.}(2013)]%
        {Anati2013InnovativeTF}
\bibfield{author}{\bibinfo{person}{Ittai Anati}, \bibinfo{person}{Shay Gueron},
  \bibinfo{person}{Simon Johnson}, {and} \bibinfo{person}{Vincent Scarlata}.}
  \bibinfo{year}{2013}\natexlab{}.
\newblock \showarticletitle{Innovative Technology for CPU Based Attestation and
  Sealing}.
\newblock
\urldef\tempurl%
\url{https://api.semanticscholar.org/CorpusID:14218854}
\showURL{%
\tempurl}


\bibitem[Anderson and Farrell(2022)]%
        {AndFar22}
\bibfield{author}{\bibinfo{person}{Connor Anderson} {and} \bibinfo{person}{Ryan
  Farrell}.} \bibinfo{year}{2022}\natexlab{}.
\newblock \showarticletitle{Improving fractal pre-training}. In
  \bibinfo{booktitle}{\emph{Proceedings of the IEEE/CVF Winter Conference on
  Applications of Computer Vision}}. \bibinfo{pages}{1300--1309}.
\newblock


\bibitem[Boenisch et~al\mbox{.}(2023)]%
        {BoeDziSchu2023when}
\bibfield{author}{\bibinfo{person}{Franziska Boenisch}, \bibinfo{person}{Adam
  Dziedzic}, \bibinfo{person}{Roei Schuster}, \bibinfo{person}{Ali~Shahin
  Shamsabadi}, \bibinfo{person}{Ilia Shumailov}, {and} \bibinfo{person}{Nicolas
  Papernot}.} \bibinfo{year}{2023}\natexlab{}.
\newblock \showarticletitle{When the Curious Abandon Honesty: Federated
  Learning Is Not Private}. In \bibinfo{booktitle}{\emph{2023 IEEE 8th European
  Symposium on Security and Privacy (EuroS\&P)}}. \bibinfo{pages}{175--199}.
\newblock
\urldef\tempurl%
\url{https://doi.org/10.1109/EuroSP57164.2023.00020}
\showDOI{\tempurl}


\bibitem[Brown et~al\mbox{.}(2010)]%
        {brown2010highly}
\bibfield{author}{\bibinfo{person}{MR Brown}, \bibinfo{person}{R Errington},
  \bibinfo{person}{Paul Rees}, \bibinfo{person}{PR Williams}, {and}
  \bibinfo{person}{SP Wilks}.} \bibinfo{year}{2010}\natexlab{}.
\newblock \showarticletitle{A highly efficient algorithm for the generation of
  random fractal aggregates}.
\newblock \bibinfo{journal}{\emph{Physica D: Nonlinear Phenomena}}
  \bibinfo{volume}{239}, \bibinfo{number}{12} (\bibinfo{year}{2010}),
  \bibinfo{pages}{1061--1066}.
\newblock


\bibitem[Canetti(2001)]%
        {Canetti01}
\bibfield{author}{\bibinfo{person}{Ran Canetti}.}
  \bibinfo{year}{2001}\natexlab{}.
\newblock \showarticletitle{Universally Composable Security: {A} New Paradigm
  for Cryptographic Protocols}. In \bibinfo{booktitle}{\emph{42nd Annual
  Symposium on Foundations of Computer Science, {FOCS}}}.
  \bibinfo{publisher}{IEEE}, \bibinfo{pages}{136--145}.
\newblock


\bibitem[Canetti(2004)]%
        {canetti2004universally}
\bibfield{author}{\bibinfo{person}{Ran Canetti}.}
  \bibinfo{year}{2004}\natexlab{}.
\newblock \showarticletitle{Universally composable signature, certification,
  and authentication}. In \bibinfo{booktitle}{\emph{Proceedings. 17th IEEE
  Computer Security Foundations Workshop, 2004.}} IEEE,
  \bibinfo{pages}{219--233}.
\newblock


\bibitem[Cao et~al\mbox{.}(2024)]%
        {Caozhazha2024SRFL}
\bibfield{author}{\bibinfo{person}{Yihao Cao}, \bibinfo{person}{Jianbiao
  Zhang}, \bibinfo{person}{Yaru Zhao}, \bibinfo{person}{Pengchong Su}, {and}
  \bibinfo{person}{Haoxiang Huang}.} \bibinfo{year}{2024}\natexlab{}.
\newblock \showarticletitle{SRFL: A Secure \& Robust Federated Learning
  framework for IoT with trusted execution environments}.
\newblock \bibinfo{journal}{\emph{Expert Systems with Applications}}
  \bibinfo{volume}{239} (\bibinfo{year}{2024}), \bibinfo{pages}{122410}.
\newblock
\showISSN{0957-4174}


\bibitem[Chase and Chow(2009)]%
        {ChaCho09}
\bibfield{author}{\bibinfo{person}{Melissa Chase} {and}
  \bibinfo{person}{Sherman~SM Chow}.} \bibinfo{year}{2009}\natexlab{}.
\newblock \showarticletitle{Improving privacy and security in multi-authority
  attribute-based encryption}. In \bibinfo{booktitle}{\emph{Proceedings of the
  16th ACM conference on Computer and communications security}}.
  \bibinfo{pages}{121--130}.
\newblock


\bibitem[Chen et~al\mbox{.}(2024)]%
        {chen2024opportunities}
\bibfield{author}{\bibinfo{person}{Minshuo Chen}, \bibinfo{person}{Song Mei},
  \bibinfo{person}{Jianqing Fan}, {and} \bibinfo{person}{Mengdi Wang}.}
  \bibinfo{year}{2024}\natexlab{}.
\newblock \showarticletitle{Opportunities and challenges of diffusion models
  for generative AI}.
\newblock \bibinfo{journal}{\emph{National Science Review}}
  \bibinfo{volume}{11}, \bibinfo{number}{12} (\bibinfo{year}{2024}),
  \bibinfo{pages}{nwae348}.
\newblock


\bibitem[Chen et~al\mbox{.}(2020)]%
        {CheNinYue20}
\bibfield{author}{\bibinfo{person}{Yujing Chen}, \bibinfo{person}{Yue Ning},
  \bibinfo{person}{Martin Slawski}, {and} \bibinfo{person}{Huzefa Rangwala}.}
  \bibinfo{year}{2020}\natexlab{}.
\newblock \showarticletitle{Asynchronous online federated learning for edge
  devices with non-iid data}. In \bibinfo{booktitle}{\emph{2020 IEEE
  International Conference on Big Data (Big Data)}}. IEEE,
  \bibinfo{pages}{15--24}.
\newblock


\bibitem[Chu and Chen(2000)]%
        {chu2000fast}
\bibfield{author}{\bibinfo{person}{Hsueh-Ting Chu} {and}
  \bibinfo{person}{Chaur-Chin Chen}.} \bibinfo{year}{2000}\natexlab{}.
\newblock \showarticletitle{A fast algorithm for generating fractals}. In
  \bibinfo{booktitle}{\emph{Proceedings 15th International Conference on
  Pattern Recognition. ICPR-2000}}, Vol.~\bibinfo{volume}{3}. IEEE,
  \bibinfo{pages}{302--305}.
\newblock


\bibitem[Corporation(2018)]%
        {intel2018ratls}
\bibfield{author}{\bibinfo{person}{Intel Corporation}.}
  \bibinfo{year}{2018}\natexlab{}.
\newblock \bibinfo{title}{Integrating Remote Attestation with Transport Layer
  Security}.
\newblock \bibinfo{howpublished}{\url{https://arxiv.org/pdf/1801.05863}}.
\newblock
\newblock
\shownote{Accessed: 2024-12-29}.


\bibitem[Dougherty et~al\mbox{.}(2021)]%
        {edgeCom}
\bibfield{author}{\bibinfo{person}{Sean Dougherty}, \bibinfo{person}{Reza
  Tourani}, \bibinfo{person}{Gaurav Panwar}, \bibinfo{person}{Roopa
  Vishwanathan}, \bibinfo{person}{Satyajayant Misra}, {and}
  \bibinfo{person}{Srikathyayani Srikanteswara}.}
  \bibinfo{year}{2021}\natexlab{}.
\newblock \showarticletitle{APECS: A Distributed Access Control Framework for
  Pervasive Edge Computing Services}. In \bibinfo{booktitle}{\emph{Proceedings
  of the 2021 ACM SIGSAC Conference on Computer and Communications Security}}
  (Virtual Event, Republic of Korea) \emph{(\bibinfo{series}{CCS '21})}.
  \bibinfo{publisher}{Association for Computing Machinery},
  \bibinfo{address}{New York, NY, USA}, \bibinfo{pages}{1405–1420}.
\newblock
\showISBNx{9781450384544}
\urldef\tempurl%
\url{https://doi.org/10.1145/3460120.3484804}
\showDOI{\tempurl}


\bibitem[Fang et~al\mbox{.}(2022)]%
        {FanLiuGon2022AFLGuard}
\bibfield{author}{\bibinfo{person}{Minghong Fang}, \bibinfo{person}{Jia Liu},
  \bibinfo{person}{Neil~Zhenqiang Gong}, {and} \bibinfo{person}{Elizabeth~S.
  Bentley}.} \bibinfo{year}{2022}\natexlab{}.
\newblock \showarticletitle{AFLGuard: Byzantine-robust Asynchronous Federated
  Learning}. In \bibinfo{booktitle}{\emph{Proceedings of the 38th Annual
  Computer Security Applications Conference}} (<conf-loc>, <city>Austin</city>,
  <state>TX</state>, <country>USA</country>, </conf-loc>)
  \emph{(\bibinfo{series}{ACSAC '22})}. \bibinfo{publisher}{Association for
  Computing Machinery}, \bibinfo{address}{New York, NY, USA},
  \bibinfo{pages}{632–646}.
\newblock
\showISBNx{9781450397599}
\urldef\tempurl%
\url{https://doi.org/10.1145/3564625.3567991}
\showDOI{\tempurl}


\bibitem[Frosst et~al\mbox{.}(2019)]%
        {FroPapHin19}
\bibfield{author}{\bibinfo{person}{Nicholas Frosst}, \bibinfo{person}{Nicolas
  Papernot}, {and} \bibinfo{person}{Geoffrey Hinton}.}
  \bibinfo{year}{2019}\natexlab{}.
\newblock \showarticletitle{Analyzing and improving representations with the
  soft nearest neighbor loss}. In \bibinfo{booktitle}{\emph{International
  conference on machine learning}}. PMLR, \bibinfo{pages}{2012--2020}.
\newblock


\bibitem[Geiping et~al\mbox{.}(2020)]%
        {invertGradiants}
\bibfield{author}{\bibinfo{person}{Jonas Geiping}, \bibinfo{person}{Hartmut
  Bauermeister}, \bibinfo{person}{Hannah Dr\"{o}ge}, {and}
  \bibinfo{person}{Michael Moeller}.} \bibinfo{year}{2020}\natexlab{}.
\newblock \showarticletitle{Inverting gradients - how easy is it to break
  privacy in federated learning?}. In \bibinfo{booktitle}{\emph{Proceedings of
  the 34th International Conference on Neural Information Processing Systems}}
  (Vancouver, BC, Canada) \emph{(\bibinfo{series}{NIPS '20})}.
  \bibinfo{publisher}{Curran Associates Inc.}, \bibinfo{address}{Red Hook, NY,
  USA}, Article \bibinfo{articleno}{1421}, \bibinfo{numpages}{11}~pages.
\newblock
\showISBNx{9781713829546}


\bibitem[He et~al\mbox{.}(2016)]%
        {resnet18}
\bibfield{author}{\bibinfo{person}{Kaiming He}, \bibinfo{person}{Xiangyu
  Zhang}, \bibinfo{person}{Shaoqing Ren}, {and} \bibinfo{person}{Jian Sun}.}
  \bibinfo{year}{2016}\natexlab{}.
\newblock \bibinfo{title}{Deep residual learning for image recognition}.
\newblock , \bibinfo{numpages}{770--778}~pages.
\newblock


\bibitem[He et~al\mbox{.}(2024)]%
        {Hewancai2024clustered}
\bibfield{author}{\bibinfo{person}{Zaobo He}, \bibinfo{person}{Lintao Wang},
  {and} \bibinfo{person}{Zhipeng Cai}.} \bibinfo{year}{2024}\natexlab{}.
\newblock \showarticletitle{Clustered Federated Learning With Adaptive Local
  Differential Privacy on Heterogeneous IoT Data}.
\newblock \bibinfo{journal}{\emph{IEEE Internet of Things Journal}}
  \bibinfo{volume}{11}, \bibinfo{number}{1} (\bibinfo{year}{2024}),
  \bibinfo{pages}{137--146}.
\newblock
\urldef\tempurl%
\url{https://doi.org/10.1109/JIOT.2023.3299947}
\showDOI{\tempurl}


\bibitem[Heged{\"u}s et~al\mbox{.}(2019)]%
        {Gossip}
\bibfield{author}{\bibinfo{person}{Istv{\'a}n Heged{\"u}s},
  \bibinfo{person}{G{\'a}bor Danner}, {and} \bibinfo{person}{M{\'a}rk
  Jelasity}.} \bibinfo{year}{2019}\natexlab{}.
\newblock \showarticletitle{Gossip Learning as a Decentralized Alternative to
  Federated Learning}. In \bibinfo{booktitle}{\emph{IFIP International
  Conference on Distributed Applications and Interoperable Systems}}.
\newblock


\bibitem[Hirokatsu et~al\mbox{.}(2022)]%
        {synthImages}
\bibfield{author}{\bibinfo{person}{Kataoka Hirokatsu},
  \bibinfo{person}{Matsumoto Asato}, \bibinfo{person}{Yamagata Eisuke},
  \bibinfo{person}{Yamada Ryosuke}, \bibinfo{person}{Inoue Nakamasa},
  \bibinfo{person}{Akio Nakamura}, {and} \bibinfo{person}{Satoh Yutaka}.}
  \bibinfo{year}{2022}\natexlab{}.
\newblock \showarticletitle{Pre-training without natural images}.
\newblock \bibinfo{journal}{\emph{International Journal of Computer Vision}}
  \bibinfo{volume}{130}, \bibinfo{number}{4} (\bibinfo{year}{2022}),
  \bibinfo{pages}{990--1007}.
\newblock


\bibitem[Hu et~al\mbox{.}(2023)]%
        {HuGuoGon2023Federated}
\bibfield{author}{\bibinfo{person}{Rui Hu}, \bibinfo{person}{Yuanxiong Guo},
  {and} \bibinfo{person}{Yanmin Gong}.} \bibinfo{year}{2023}\natexlab{}.
\newblock \showarticletitle{Federated Learning with Sparsified Model
  Perturbation: Improving Accuracy under Client-Level Differential Privacy}.
\newblock \bibinfo{journal}{\emph{IEEE Transactions on Mobile Computing}}
  (\bibinfo{year}{2023}), \bibinfo{pages}{1--14}.
\newblock
\urldef\tempurl%
\url{https://doi.org/10.1109/TMC.2023.3343288}
\showDOI{\tempurl}


\bibitem[Huang et~al\mbox{.}(2008)]%
        {Hua08}
\bibfield{author}{\bibinfo{person}{Anna Huang} {et~al\mbox{.}}}
  \bibinfo{year}{2008}\natexlab{}.
\newblock \showarticletitle{Similarity measures for text document clustering}.
  In \bibinfo{booktitle}{\emph{Proceedings of the sixth new zealand computer
  science research student conference (NZCSRSC2008), Christchurch, New
  Zealand}}, Vol.~\bibinfo{volume}{4}. \bibinfo{pages}{9--56}.
\newblock


\bibitem[Jia et~al\mbox{.}(2021)]%
        {JiaChoCha21}
\bibfield{author}{\bibinfo{person}{Hengrui Jia},
  \bibinfo{person}{Christopher~A. Choquette-Choo}, \bibinfo{person}{Varun
  Chandrasekaran}, {and} \bibinfo{person}{Nicolas Papernot}.}
  \bibinfo{year}{2021}\natexlab{}.
\newblock \showarticletitle{Entangled Watermarks as a Defense against Model
  Extraction}. In \bibinfo{booktitle}{\emph{30th USENIX Security Symposium
  (USENIX Security 21)}}. \bibinfo{publisher}{USENIX Association},
  \bibinfo{pages}{1937--1954}.
\newblock
\showISBNx{978-1-939133-24-3}
\urldef\tempurl%
\url{https://www.usenix.org/conference/usenixsecurity21/presentation/jia}
\showURL{%
\tempurl}


\bibitem[Kataoka et~al\mbox{.}(2020)]%
        {KatOkaKaz20}
\bibfield{author}{\bibinfo{person}{Hirokatsu Kataoka},
  \bibinfo{person}{Kazushige Okayasu}, \bibinfo{person}{Asato Matsumoto},
  \bibinfo{person}{Eisuke Yamagata}, \bibinfo{person}{Ryosuke Yamada},
  \bibinfo{person}{Nakamasa Inoue}, \bibinfo{person}{Akio Nakamura}, {and}
  \bibinfo{person}{Yutaka Satoh}.} \bibinfo{year}{2020}\natexlab{}.
\newblock \showarticletitle{Pre-training without natural images}. In
  \bibinfo{booktitle}{\emph{Proceedings of the Asian Conference on Computer
  Vision}}.
\newblock


\bibitem[Konecn{\'y} et~al\mbox{.}(2016)]%
        {FederatedOpt}
\bibfield{author}{\bibinfo{person}{Jakub Konecn{\'y}}, \bibinfo{person}{H.~B.
  McMahan}, \bibinfo{person}{Daniel Ramage}, {and} \bibinfo{person}{Peter
  Richt{\'a}rik}.} \bibinfo{year}{2016}\natexlab{}.
\newblock \showarticletitle{Federated Optimization: Distributed Machine
  Learning for On-Device Intelligence}.
\newblock \bibinfo{journal}{\emph{ArXiv}}  \bibinfo{volume}{abs/1610.02527}
  (\bibinfo{year}{2016}).
\newblock


\bibitem[Krizhevsky(2012)]%
        {CIFAR}
\bibfield{author}{\bibinfo{person}{Alex Krizhevsky}.}
  \bibinfo{year}{2012}\natexlab{}.
\newblock \showarticletitle{Learning Multiple Layers of Features from Tiny
  Images}.
\newblock \bibinfo{journal}{\emph{University of Toronto}} (\bibinfo{date}{05}
  \bibinfo{year}{2012}).
\newblock


\bibitem[Liu et~al\mbox{.}(2023)]%
        {liuligao2023privacy-encoded}
\bibfield{author}{\bibinfo{person}{Hongfu Liu}, \bibinfo{person}{Bin Li},
  \bibinfo{person}{Changlong Gao}, \bibinfo{person}{Pei Xie}, {and}
  \bibinfo{person}{Chenglin Zhao}.} \bibinfo{year}{2023}\natexlab{}.
\newblock \showarticletitle{Privacy-Encoded Federated Learning Against
  Gradient-Based Data Reconstruction Attacks}.
\newblock \bibinfo{journal}{\emph{IEEE Transactions on Information Forensics
  and Security}}  \bibinfo{volume}{18} (\bibinfo{year}{2023}),
  \bibinfo{pages}{5860--5875}.
\newblock
\urldef\tempurl%
\url{https://doi.org/10.1109/TIFS.2023.3309095}
\showDOI{\tempurl}


\bibitem[Liu et~al\mbox{.}(2024)]%
        {liuyuzon2024Delay}
\bibfield{author}{\bibinfo{person}{Shumei Liu}, \bibinfo{person}{Yao Yu},
  \bibinfo{person}{Yue Zong}, \bibinfo{person}{Phee~Lep Yeoh},
  \bibinfo{person}{Lei Guo}, \bibinfo{person}{Branka Vucetic},
  \bibinfo{person}{Trung~Q. Duong}, {and} \bibinfo{person}{Yonghui Li}.}
  \bibinfo{year}{2024}\natexlab{}.
\newblock \showarticletitle{Delay and Energy-Efficient Asynchronous Federated
  Learning for Intrusion Detection in Heterogeneous Industrial Internet of
  Things}.
\newblock \bibinfo{journal}{\emph{IEEE Internet of Things Journal}}
  \bibinfo{volume}{11}, \bibinfo{number}{8} (\bibinfo{year}{2024}),
  \bibinfo{pages}{14739--14754}.
\newblock
\urldef\tempurl%
\url{https://doi.org/10.1109/JIOT.2023.3344457}
\showDOI{\tempurl}


\bibitem[Luo et~al\mbox{.}(2018)]%
        {ChuJiaLei17}
\bibfield{author}{\bibinfo{person}{Chunjie Luo}, \bibinfo{person}{Jianfeng
  Zhan}, \bibinfo{person}{Xiaohe Xue}, \bibinfo{person}{Lei Wang},
  \bibinfo{person}{Rui Ren}, {and} \bibinfo{person}{Qiang Yang}.}
  \bibinfo{year}{2018}\natexlab{}.
\newblock \bibinfo{title}{Cosine normalization: Using cosine similarity instead
  of dot product in neural networks}.
\newblock , \bibinfo{numpages}{382--391}~pages.
\newblock


\bibitem[Mammen(2021)]%
        {Mam21}
\bibfield{author}{\bibinfo{person}{Priyanka~Mary Mammen}.}
  \bibinfo{year}{2021}\natexlab{}.
\newblock \showarticletitle{Federated learning: Opportunities and challenges}.
\newblock \bibinfo{journal}{\emph{arXiv preprint arXiv:2101.05428}}
  (\bibinfo{year}{2021}).
\newblock


\bibitem[McKeen et~al\mbox{.}(2013)]%
        {IEECiteforSGX}
\bibfield{author}{\bibinfo{person}{Francis~X. McKeen}, \bibinfo{person}{Ilya
  Alexandrovich}, \bibinfo{person}{Alex Berenzon}, \bibinfo{person}{Carlos~V.
  Rozas}, \bibinfo{person}{Hisham Shafi}, \bibinfo{person}{Vedvyas Shanbhogue},
  {and} \bibinfo{person}{Uday~R. Savagaonkar}.}
  \bibinfo{year}{2013}\natexlab{}.
\newblock \showarticletitle{Innovative instructions and software model for
  isolated execution}. In \bibinfo{booktitle}{\emph{Hardware and Architectural
  Support for Security and Privacy}}.
\newblock
\urldef\tempurl%
\url{https://api.semanticscholar.org/CorpusID:40428970}
\showURL{%
\tempurl}


\bibitem[Melis et~al\mbox{.}(2019)]%
        {melsondec2019exploiting}
\bibfield{author}{\bibinfo{person}{Luca Melis}, \bibinfo{person}{Congzheng
  Song}, \bibinfo{person}{Emiliano De~Cristofaro}, {and}
  \bibinfo{person}{Vitaly Shmatikov}.} \bibinfo{year}{2019}\natexlab{}.
\newblock \showarticletitle{Exploiting Unintended Feature Leakage in
  Collaborative Learning}. In \bibinfo{booktitle}{\emph{2019 IEEE Symposium on
  Security and Privacy (SP)}}. \bibinfo{pages}{691--706}.
\newblock
\urldef\tempurl%
\url{https://doi.org/10.1109/SP.2019.00029}
\showDOI{\tempurl}


\bibitem[Na et~al\mbox{.}(2022)]%
        {NaHyeJun2022Closing}
\bibfield{author}{\bibinfo{person}{Seung~Ho Na}, \bibinfo{person}{Hyeong~Gwon
  Hong}, \bibinfo{person}{Junmo Kim}, {and} \bibinfo{person}{Seungwon Shin}.}
  \bibinfo{year}{2022}\natexlab{}.
\newblock \showarticletitle{Closing the Loophole: Rethinking Reconstruction
  Attacks in Federated Learning from a Privacy Standpoint}. In
  \bibinfo{booktitle}{\emph{Proceedings of the 38th Annual Computer Security
  Applications Conference}} (<conf-loc>, <city>Austin</city>,
  <state>TX</state>, <country>USA</country>, </conf-loc>)
  \emph{(\bibinfo{series}{ACSAC '22})}. \bibinfo{publisher}{Association for
  Computing Machinery}, \bibinfo{address}{New York, NY, USA},
  \bibinfo{pages}{332–345}.
\newblock
\showISBNx{9781450397599}
\urldef\tempurl%
\url{https://doi.org/10.1145/3564625.3564657}
\showDOI{\tempurl}


\bibitem[Nasr et~al\mbox{.}(2019)]%
        {nasshohou2019comprehensive}
\bibfield{author}{\bibinfo{person}{Milad Nasr}, \bibinfo{person}{Reza Shokri},
  {and} \bibinfo{person}{Amir Houmansadr}.} \bibinfo{year}{2019}\natexlab{}.
\newblock \showarticletitle{Comprehensive Privacy Analysis of Deep Learning:
  Passive and Active White-box Inference Attacks against Centralized and
  Federated Learning}. In \bibinfo{booktitle}{\emph{2019 IEEE Symposium on
  Security and Privacy (SP)}}. \bibinfo{pages}{739--753}.
\newblock
\urldef\tempurl%
\url{https://doi.org/10.1109/SP.2019.00065}
\showDOI{\tempurl}


\bibitem[Nguyen et~al\mbox{.}(2022)]%
        {ngumalzha2022federated}
\bibfield{author}{\bibinfo{person}{John Nguyen}, \bibinfo{person}{Kshitiz
  Malik}, \bibinfo{person}{Hongyuan Zhan}, \bibinfo{person}{Ashkan Yousefpour},
  \bibinfo{person}{Mike Rabbat}, \bibinfo{person}{Mani Malek}, {and}
  \bibinfo{person}{Dzmitry Huba}.} \bibinfo{year}{2022}\natexlab{}.
\newblock \showarticletitle{Federated learning with buffered asynchronous
  aggregation}. In \bibinfo{booktitle}{\emph{International Conference on
  Artificial Intelligence and Statistics}}. PMLR, \bibinfo{pages}{3581--3607}.
\newblock


\bibitem[Pasquini et~al\mbox{.}(2022)]%
        {pasfraate2022Eluding}
\bibfield{author}{\bibinfo{person}{Dario Pasquini}, \bibinfo{person}{Danilo
  Francati}, {and} \bibinfo{person}{Giuseppe Ateniese}.}
  \bibinfo{year}{2022}\natexlab{}.
\newblock \showarticletitle{Eluding Secure Aggregation in Federated Learning
  via Model Inconsistency}. In \bibinfo{booktitle}{\emph{Proceedings of the
  2022 ACM SIGSAC Conference on Computer and Communications Security}} (Los
  Angeles, CA, USA) \emph{(\bibinfo{series}{CCS '22})}.
  \bibinfo{publisher}{Association for Computing Machinery},
  \bibinfo{address}{New York, NY, USA}, \bibinfo{pages}{2429–2443}.
\newblock
\showISBNx{9781450394505}
\urldef\tempurl%
\url{https://doi.org/10.1145/3548606.3560557}
\showDOI{\tempurl}


\bibitem[Pass et~al\mbox{.}(2017)]%
        {pass2017formal}
\bibfield{author}{\bibinfo{person}{Rafael Pass}, \bibinfo{person}{Elaine Shi},
  {and} \bibinfo{person}{Florian Tramer}.} \bibinfo{year}{2017}\natexlab{}.
\newblock \showarticletitle{Formal abstractions for attested execution secure
  processors}. In \bibinfo{booktitle}{\emph{Advances in Cryptology--EUROCRYPT
  2017: 36th Annual International Conference on the Theory and Applications of
  Cryptographic Techniques, Paris, France, April 30--May 4, 2017, Proceedings,
  Part I 36}}. Springer, \bibinfo{pages}{260--289}.
\newblock


\bibitem[Pichler et~al\mbox{.}(2023)]%
        {picromveg2023Perfectly}
\bibfield{author}{\bibinfo{person}{Georg Pichler}, \bibinfo{person}{Marco
  Romanelli}, \bibinfo{person}{Leonardo~Rey Vega}, {and} \bibinfo{person}{Pablo
  Piantanida}.} \bibinfo{year}{2023}\natexlab{}.
\newblock \showarticletitle{Perfectly Accurate Membership Inference by a
  Dishonest Central Server in Federated Learning}.
\newblock \bibinfo{journal}{\emph{IEEE Transactions on Dependable and Secure
  Computing}} (\bibinfo{year}{2023}), \bibinfo{pages}{1--8}.
\newblock
\urldef\tempurl%
\url{https://doi.org/10.1109/TDSC.2023.3326230}
\showDOI{\tempurl}


\bibitem[Project(2021)]%
        {gramine}
\bibfield{author}{\bibinfo{person}{Gramine Project}.}
  \bibinfo{year}{2021}\natexlab{}.
\newblock \bibinfo{title}{Gramine Library OS with Intel SGX Support}.
\newblock \bibinfo{howpublished}{\url{https://gramineproject.io/}}.
\newblock
\newblock
\shownote{Accessed: 2024-12-29}.


\bibitem[Sabt et~al\mbox{.}(2015)]%
        {TEECite}
\bibfield{author}{\bibinfo{person}{Mohamed Sabt}, \bibinfo{person}{Mohammed
  Achemlal}, {and} \bibinfo{person}{Abdelmadjid Bouabdallah}.}
  \bibinfo{year}{2015}\natexlab{}.
\newblock \showarticletitle{Trusted Execution Environment: What It is, and What
  It is Not}. In \bibinfo{booktitle}{\emph{2015 IEEE Trustcom/BigDataSE/ISPA}},
  Vol.~\bibinfo{volume}{1}. \bibinfo{pages}{57--64}.
\newblock
\urldef\tempurl%
\url{https://doi.org/10.1109/Trustcom.2015.357}
\showDOI{\tempurl}


\bibitem[Shokri and Shmatikov(2015)]%
        {ShoRezShm2015Privacy}
\bibfield{author}{\bibinfo{person}{Reza Shokri} {and} \bibinfo{person}{Vitaly
  Shmatikov}.} \bibinfo{year}{2015}\natexlab{}.
\newblock \showarticletitle{Privacy-Preserving Deep Learning}. In
  \bibinfo{booktitle}{\emph{Proceedings of the 22nd ACM SIGSAC Conference on
  Computer and Communications Security}} (Denver, Colorado, USA)
  \emph{(\bibinfo{series}{CCS '15})}. \bibinfo{publisher}{Association for
  Computing Machinery}, \bibinfo{address}{New York, NY, USA},
  \bibinfo{pages}{1310–1321}.
\newblock
\showISBNx{9781450338325}
\urldef\tempurl%
\url{https://doi.org/10.1145/2810103.2813687}
\showDOI{\tempurl}


\bibitem[Shokri et~al\mbox{.}(2017)]%
        {shostrmarc2017membership}
\bibfield{author}{\bibinfo{person}{Reza Shokri}, \bibinfo{person}{Marco
  Stronati}, \bibinfo{person}{Congzheng Song}, {and} \bibinfo{person}{Vitaly
  Shmatikov}.} \bibinfo{year}{2017}\natexlab{}.
\newblock \showarticletitle{Membership Inference Attacks Against Machine
  Learning Models}. In \bibinfo{booktitle}{\emph{2017 IEEE Symposium on
  Security and Privacy (SP)}}. \bibinfo{pages}{3--18}.
\newblock
\urldef\tempurl%
\url{https://doi.org/10.1109/SP.2017.41}
\showDOI{\tempurl}


\bibitem[Stallkamp et~al\mbox{.}(2012)]%
        {stallkamp2012man}
\bibfield{author}{\bibinfo{person}{Johannes Stallkamp}, \bibinfo{person}{Marc
  Schlipsing}, \bibinfo{person}{Jan Salmen}, {and} \bibinfo{person}{Christian
  Igel}.} \bibinfo{year}{2012}\natexlab{}.
\newblock \showarticletitle{Man vs. Computer: Benchmarking Machine Learning
  Algorithms for Traffic Sign Recognition}.
\newblock \bibinfo{journal}{\emph{Neural Networks}}  \bibinfo{volume}{32}
  (\bibinfo{year}{2012}), \bibinfo{pages}{323--332}.
\newblock
\urldef\tempurl%
\url{https://doi.org/10.1016/j.neunet.2012.02.016}
\showDOI{\tempurl}


\bibitem[Tian et~al\mbox{.}(2021)]%
        {tiacheyu2021towards}
\bibfield{author}{\bibinfo{person}{Pu Tian}, \bibinfo{person}{Zheyi Chen},
  \bibinfo{person}{Wei Yu}, {and} \bibinfo{person}{Weixian Liao}.}
  \bibinfo{year}{2021}\natexlab{}.
\newblock \showarticletitle{Towards asynchronous federated learning based
  threat detection: A DC-Adam approach}.
\newblock \bibinfo{journal}{\emph{Computers \& Security}}
  \bibinfo{volume}{108} (\bibinfo{year}{2021}), \bibinfo{pages}{102344}.
\newblock
\showISSN{0167-4048}
\urldef\tempurl%
\url{https://doi.org/10.1016/j.cose.2021.102344}
\showDOI{\tempurl}


\bibitem[Ulhaq and Akhtar(2022)]%
        {ulhaq2022efficient}
\bibfield{author}{\bibinfo{person}{Anwaar Ulhaq} {and} \bibinfo{person}{Naveed
  Akhtar}.} \bibinfo{year}{2022}\natexlab{}.
\newblock \showarticletitle{Efficient diffusion models for vision: A survey}.
\newblock \bibinfo{journal}{\emph{arXiv preprint arXiv:2210.09292}}
  (\bibinfo{year}{2022}).
\newblock


\bibitem[Wei et~al\mbox{.}(2023)]%
        {weilima2023personalized}
\bibfield{author}{\bibinfo{person}{Kang Wei}, \bibinfo{person}{Jun Li},
  \bibinfo{person}{Chuan Ma}, \bibinfo{person}{Ming Ding}, \bibinfo{person}{Wen
  Chen}, \bibinfo{person}{Jun Wu}, \bibinfo{person}{Meixia Tao}, {and}
  \bibinfo{person}{H.~Vincent Poor}.} \bibinfo{year}{2023}\natexlab{}.
\newblock \showarticletitle{Personalized Federated Learning With Differential
  Privacy and Convergence Guarantee}.
\newblock \bibinfo{journal}{\emph{IEEE Transactions on Information Forensics
  and Security}}  \bibinfo{volume}{18} (\bibinfo{year}{2023}),
  \bibinfo{pages}{4488--4503}.
\newblock
\urldef\tempurl%
\url{https://doi.org/10.1109/TIFS.2023.3293417}
\showDOI{\tempurl}


\bibitem[Xiao et~al\mbox{.}(2017)]%
        {FMNIST}
\bibfield{author}{\bibinfo{person}{Han Xiao}, \bibinfo{person}{Kashif Rasul},
  {and} \bibinfo{person}{Roland Vollgraf}.} \bibinfo{year}{2017}\natexlab{}.
\newblock \bibinfo{booktitle}{\emph{Fashion-MNIST: a Novel Image Dataset for
  Benchmarking Machine Learning Algorithms}}.
\newblock
\showeprint[arXiv]{cs.LG/1708.07747}~[cs.LG]


\bibitem[Xu et~al\mbox{.}(2020)]%
        {xuliliu2020verifynet}
\bibfield{author}{\bibinfo{person}{Guowen Xu}, \bibinfo{person}{Hongwei Li},
  \bibinfo{person}{Sen Liu}, \bibinfo{person}{Kan Yang}, {and}
  \bibinfo{person}{Xiaodong Lin}.} \bibinfo{year}{2020}\natexlab{}.
\newblock \showarticletitle{VerifyNet: Secure and Verifiable Federated
  Learning}.
\newblock \bibinfo{journal}{\emph{IEEE Transactions on Information Forensics
  and Security}}  \bibinfo{volume}{15} (\bibinfo{year}{2020}),
  \bibinfo{pages}{911--926}.
\newblock
\urldef\tempurl%
\url{https://doi.org/10.1109/TIFS.2019.2929409}
\showDOI{\tempurl}


\bibitem[Yang et~al\mbox{.}(2023)]%
        {yanGeXia2023using}
\bibfield{author}{\bibinfo{person}{Haomiao Yang}, \bibinfo{person}{Mengyu Ge},
  \bibinfo{person}{Kunlan Xiang}, {and} \bibinfo{person}{Jingwei Li}.}
  \bibinfo{year}{2023}\natexlab{}.
\newblock \showarticletitle{Using Highly Compressed Gradients in Federated
  Learning for Data Reconstruction Attacks}.
\newblock \bibinfo{journal}{\emph{IEEE Transactions on Information Forensics
  and Security}}  \bibinfo{volume}{18} (\bibinfo{year}{2023}),
  \bibinfo{pages}{818--830}.
\newblock
\urldef\tempurl%
\url{https://doi.org/10.1109/TIFS.2022.3227761}
\showDOI{\tempurl}


\bibitem[Yin et~al\mbox{.}(2021)]%
        {batchRecovery}
\bibfield{author}{\bibinfo{person}{Hongxu Yin}, \bibinfo{person}{Arun Mallya},
  \bibinfo{person}{Arash Vahdat}, \bibinfo{person}{Jose~M. Alvarez},
  \bibinfo{person}{Jan Kautz}, {and} \bibinfo{person}{Pavlo Molchanov}.}
  \bibinfo{year}{2021}\natexlab{}.
\newblock \showarticletitle{See through Gradients: Image Batch Recovery via
  GradInversion}. In \bibinfo{booktitle}{\emph{2021 IEEE/CVF Conference on
  Computer Vision and Pattern Recognition (CVPR)}}.
  \bibinfo{pages}{16332--16341}.
\newblock
\urldef\tempurl%
\url{https://doi.org/10.1109/CVPR46437.2021.01607}
\showDOI{\tempurl}


\bibitem[Yue et~al\mbox{.}(2023)]%
        {gradientObfuscation}
\bibfield{author}{\bibinfo{person}{Kai Yue}, \bibinfo{person}{Richeng Jin},
  \bibinfo{person}{Chau-Wai Wong}, \bibinfo{person}{Dror Baron}, {and}
  \bibinfo{person}{Huaiyu Dai}.} \bibinfo{year}{2023}\natexlab{}.
\newblock \showarticletitle{Gradient obfuscation gives a false sense of
  security in federated learning}. In \bibinfo{booktitle}{\emph{Proceedings of
  the 32nd USENIX Conference on Security Symposium}} (Anaheim, CA, USA)
  \emph{(\bibinfo{series}{SEC '23})}. \bibinfo{publisher}{USENIX Association},
  \bibinfo{address}{USA}, Article \bibinfo{articleno}{357},
  \bibinfo{numpages}{18}~pages.
\newblock
\showISBNx{978-1-939133-37-3}


\bibitem[Zhang et~al\mbox{.}(2017)]%
        {UTKFace}
\bibfield{author}{\bibinfo{person}{Zhifei Zhang}, \bibinfo{person}{Yang Song},
  \bibinfo{person}{}, {and} \bibinfo{person}{Hairong Qi}.}
  \bibinfo{year}{2017}\natexlab{}.
\newblock \showarticletitle{Age Progression/Regression by Conditional
  Adversarial Autoencoder}. In \bibinfo{booktitle}{\emph{IEEE Conference on
  Computer Vision and Pattern Recognition (CVPR)}}. IEEE.
\newblock


\bibitem[Zhao et~al\mbox{.}(2024)]%
        {ZhaShaElk2024Large-scale}
\bibfield{author}{\bibinfo{person}{J. Zhao}, \bibinfo{person}{A. Sharma},
  \bibinfo{person}{A. Elkordy}, \bibinfo{person}{Y.~H. Ezzeldin},
  \bibinfo{person}{S. Avestimehr}, {and} \bibinfo{person}{S. Bagchi}.}
  \bibinfo{year}{2024}\natexlab{}.
\newblock \showarticletitle{LOKI: Large-scale Data Reconstruction Attack
  against Federated Learning through Model Manipulation}. In
  \bibinfo{booktitle}{\emph{2024 IEEE Symposium on Security and Privacy (SP)}}.
  \bibinfo{publisher}{IEEE Computer Society}, \bibinfo{address}{Los Alamitos,
  CA, USA}, \bibinfo{pages}{34--34}.
\newblock
\showISSN{2375-1207}
\urldef\tempurl%
\url{https://doi.org/10.1109/SP54263.2024.00030}
\showDOI{\tempurl}


\bibitem[Zhu et~al\mbox{.}(2024)]%
        {zhuLiGu2024evaluating}
\bibfield{author}{\bibinfo{person}{Gongxi Zhu}, \bibinfo{person}{Donghao Li},
  \bibinfo{person}{Hanlin Gu}, \bibinfo{person}{Yuxing Han},
  \bibinfo{person}{Yuan Yao}, \bibinfo{person}{Lixin Fan}, {and}
  \bibinfo{person}{Qiang Yang}.} \bibinfo{year}{2024}\natexlab{}.
\newblock \showarticletitle{Evaluating Membership Inference Attacks and
  Defenses in Federated Learning}.
\newblock \bibinfo{journal}{\emph{arXiv preprint arXiv:2402.06289}}
  (\bibinfo{year}{2024}).
\newblock


\bibitem[Zhu et~al\mbox{.}(2019)]%
        {deepLeakage}
\bibfield{author}{\bibinfo{person}{Ligeng Zhu}, \bibinfo{person}{Zhijian Liu},
  {and} \bibinfo{person}{Song Han}.} \bibinfo{year}{2019}\natexlab{}.
\newblock \bibinfo{booktitle}{\emph{Deep leakage from gradients}}.
\newblock \bibinfo{publisher}{Curran Associates Inc.}, \bibinfo{address}{Red
  Hook, NY, USA}.
\newblock


\end{thebibliography}
